\documentclass[aps,prl,twocolumn,preprintnumbers,superscriptaddress,10pt,footinbib]{revtex4-2}
\pdfoutput=1 %

\usepackage[mathscr]{eucal}
\usepackage[dvipsnames]{xcolor}

\usepackage{epsfig,latexsym,amsfonts,amsmath,amsthm,amssymb,amsbsy,multirow,slashed,wasysym,textcomp,subfigure,hyperref,wrapfig,datetime,comment,mathtools,cancel}

\newcommand{\badat}{\begin{alignedat}}
\newcommand{\eadat}{\end{alignedat}}

\newcommand{\cA}{{\bar A}}
\newcommand{\cC}{{\bar C}}

\begin{document}

\title{Asymptotic Symmetries for Logarithmic Soft Theorems in Gauge Theory and Gravity}
\author{Sangmin Choi}
\email{s.choi@uva.nl}
\affiliation{Institute for Theoretical Physics, University of Amsterdam, PO Box 94485, 1090 GL Amsterdam, The Netherlands}
\affiliation{CPHT, CNRS, Ecole polytechnique, IP Paris, F-91128 Palaiseau, France}

\author{Alok Laddha}
\email{aladdha@cmi.ac.in}
\affiliation{Chennai Mathematical Institute, H1, SIPCOT IT Park, Siruseri, Kelambakkam 603103, India}

\author{Andrea Puhm}
\email{a.puhm@uva.nl}
\affiliation{Institute for Theoretical Physics, University of Amsterdam, PO Box 94485, 1090 GL Amsterdam, The Netherlands}
\affiliation{CPHT, CNRS, Ecole polytechnique, IP Paris, F-91128 Palaiseau, France}

\begin{abstract}
  \noindent
Gauge theories and perturbative gravity in four dimensions are governed by a tower of infinite-dimensional symmetries which arise from tree-level soft theorems.  However, aside from the leading soft theorems which are all-loop exact, subleading ones receive loop corrections due to long-range infrared effects which result in new soft theorems with logarithmic dependence on the energy of the soft particle.
The conjectured universality of these logarithmic soft theorems to all loop orders cries out for a symmetry interpretation.
In this letter we initiate a program to compute long-range infrared corrections to the charges that generate the asymptotic symmetries in (scalar) QED and perturbative gravity.
For late-time fall-offs of the electromagnetic and gravitational fields which give rise to infrared dressings for the matter fields, we derive finite charge conservation laws and show that in the quantum theory they correspond precisely to the first among the infinite tower of logarithmic soft theorems. This symmetry interpretation, by virtue of being universal and all-loop exact, is a key element for a holographic principle in spacetimes with flat asymptotics.
\end{abstract}

\maketitle
\section{Introduction}

Soft factorization theorems in gauge theories and gravity have received renewed attention over the past decade due to their deep connection with asymptotic symmetries and the quest for a holographic principle for quantum gravity in spacetimes with flat asymptotics. Indeed the knowledge of universal properties of the S-matrix, which is the natural observable in flat space, is bound to impose powerful constraints on any flat space holographic proposal. 
Tree-level amplitudes in QED (or gravity) of $N$ particles with {\it hard} momenta $p_{1},\, \dots,\, p_{N}$ and a photon (or a graviton) of {\it soft} momentum $p_s^{\mu}=\omega q^\mu(\theta,\phi)$ and helicity~$\ell$, admit a soft (i.e.\ low-energy) expansion \cite{Low:1958sn,Weinberg:1965nx,Cachazo:2014fwa,Hamada:2018vrw,Li:2018gnc} 
\begin{align}\label{treesoft}
    &{\cal M}_{N+1}(p_{1}, \dots, p_{N}; (\omega,q,\ell))
    \nonumber\\&=
    \sum_{n=-1}^{\infty} \omega^{n}S_{n}(\{p_{i}\}, (q,\ell)) {\cal M}_{N}(p_{1}, \dots, p_{N})+\dots,
\end{align}
where $S_{-1}$ is the Weinberg (leading) soft factor and $S_{n\geq 0}$ is the sub$^{n+1}$-leading soft photon (or graviton) factor.
Acting with the differential operator $\lim_{\omega \rightarrow 0} \partial_{\omega}^{n+1} \omega$ on both sides of \eqref{treesoft} extracts the sub$^{n+1}$-leading soft theorem. 
The leading soft theorem ($n=-1$) is tree-level exact (no loop corrections) and universal (theory-independent), i.e.\ $S_{-1}$ is uniquely fixed by the charges and momenta of the scattering states in QED and gravity but is independent of the details of the hard amplitude. The subleading soft theorem ($n=0$) is universal only in gravity with $S_0$ fixed by the momenta and angular momenta of the scattering states, while the form of all other (tree-level) soft theorems is theory-dependent
\cite{Elvang:2016qvq,Ghosh:2021bam,Laddha:2017ygw,Krishna:2023fxg} as indicated by the $\dots$ in \eqref{treesoft}. Factorization at subleading orders only holds for a ``projected amplitude'' where the $\dots$ do not contribute \cite{Li:2018gnc}.

The primary reason for the resurgence of interest in soft theorems are the seminal results by Strominger and collaborators \cite{Strominger:2013jfa,He:2014laa,Cachazo:2014fwa,Kapec:2014opa,Lysov:2014csa,He:2014cra,Kapec:2015ena,Campiglia:2015qka,Campiglia:2016hvg,Campiglia:2015kxa,Campiglia:2014yka} which identified the leading soft photon and graviton theorems and the subleading soft graviton theorem 
as Ward identities associated to an algebra of infinite-dimensional asymptotic symmetries: BMS supertranslations and superrotations in gravity and superphaserotations in QED \footnote{Asymptotic symmetries in gauge theory are generically called large gauge transformations. In analogy to gravity, we refer to the leading one as superphaserotation.}.
In fact, we now know that if all the external particles are massless, the entire tree-level soft expansion \eqref{treesoft} after projection to the universal part (see \cite{Li:2018gnc}) corresponds to Ward identities for a tower of asymptotic symmetries of the S-matrix \cite{Campiglia:2018dyi,Freidel:2021ytz}. %

If the charged states are massive, then the connection between asymptotic symmetries and soft theorems is more nuanced as the scattering states are localized on the union of null {\it and} time-like infinity. It has been proven that the leading and subleading soft theorems are equivalent to the Ward identities of the leading and subleading asymptotic symmetries \cite{Lysov:2014csa,He:2014cra,Kapec:2015ena,Campiglia:2015qka,Campiglia:2016hvg}, but it has yet to be shown that this symmetry interpretation persists to all sub$^{n+1}$-leading orders in the soft expansion. 

Identifying all the symmetries of gauge theories and quantum gravity in asymptotically flat spacetimes is central to establishing a flat space holographic principle. It has been a driving force behind the celestial holography program \cite{Raclariu:2021zjz,Prema:2021sjp,Pasterski:2021rjz,McLoughlin:2022ljp,Donnay:2023mrd} where, by mapping the S-matrix to correlation functions of operators in a boost basis \cite{deBoer:2003vf,Cheung:2016iub,Pasterski:2016qvg,Pasterski:2017kqt,Pasterski:2017ylz}, each term in the soft expansion with power $\omega^n$ takes the form of a Ward identity for a soft operator $\mathcal O_\Delta$ with boost weight $\Delta=-n$ in a ``celestial CFT'' living on the celestial sphere \cite{He:2014laa,Kapec:2016jld,Nande:2017dba,Donnay:2018neh,Fan:2019emx,Nandan:2019jas,Pate:2019mfs,Adamo:2019ipt,Puhm:2019zbl,Guevara:2019ypd,Kapec:2017gsg,Donnay:2020guq,Kapec:2021eug,Pasterski:2021fjn,Donnay:2022sdg,Pano:2023slc}.
The state of the art in $D=4$ spacetime dimensions is that the infinite tower of soft operators extracted from the tree-level soft theorems in \eqref{treesoft} obeys an infinite-dimensional algebra~\cite{Guevara:2021abz,Strominger:2021mtt,Freidel:2021ytz,
Geiller:2024bgf}%
; in gravity this is the $w_{1+\infty}$ algebra that appeared already in Penrose's twistor construction~\cite{Penrose:1976js}. 

The fate of this infinite-dimensional algebra at loop-level remains an open question. Indeed, only the leading soft theorem in QED and gravity is all-loop exact \footnote{Note that in $D>4$ dimensions these soft theorems are all-loop exact.} while sub$^{n+1}$-leading soft theorems are not ``infrared-stable'' at loop-level. The fact that the $D=4$ Dyson S-matrix does not exist further complicates the situation. An attempt to simply regulate the virtual infrared divergences of the S-matrix leads to (loop-corrected) soft theorems, which are however scheme-dependent \cite{Bern:2014oka}.

In a series of seminal papers, Sahoo and Sen as well as Saha, Sahoo and Sen proved a number of striking results for classical and quantum scattering
that showed that long-range infrared effects in four dimensions lead to novel soft theorems that resolve the aforementioned issues
\cite{Sahoo:2018lxl,Saha:2019tub,Sahoo:2020ryf}. 
Given the existence of a well-defined soft expansion in QED and gravity to all orders in the loop expansion, 
they showed that the ratio between an $(N+1)$-point amplitude with a soft photon (or graviton) and the $N$-point amplitude without that soft gauge boson is infrared-finite.
Its soft expansion is given by
\begin{align}\label{logsoftexp}
    &\frac{{\cal M}_{N+1}(p_{1}, \dots, p_{N};(\omega,q,\ell))}{{\cal M}_{N}(p_{1}, \dots, p_{N})}\nonumber \\ 
    &\quad\quad\quad=
    \sum_{n=-1}^{\infty} \omega^{n} (\ln\omega)^{n+1} {S}^{(\ln \omega)}_{n}+\dots
\end{align}
It starts with the $n=-1$ leading (Weinberg) soft factor ${S}^{(\ln \omega)}_{-1}\equiv {S}_{-1}$ but differs for $n\ge0$ from the tree-level soft expansion.
In \cite{Sahoo:2018lxl} the leading logarithmic soft factor $S^{(\ln \omega)}_0$ was shown to be universal and uniquely fixed by the charges and the momenta of the scattering states, and in \cite{Sahoo:2020ryf} this was shown to be true also for $S^{(\ln \omega)}_1$. 
It is conjectured by Sahoo-Sen \cite{Saha:2019tub} 
that the all-order soft expansion contains a universal tower of soft factors $S^{(\ln \omega)}_n$ (the $\dots$ contain non-universal terms of the form $\omega^n (\ln \omega)^m$ with $m\neq n+1$).
This would mean that the logarithmic soft theorems are valid to all orders in perturbation theory and independent of the details of the hard scattering!

This presents us with the truly remarkable prospect: 
{\it Is the universality of the all-loop exact logarithmic soft theorems a consequence of asymptotic symmetries of the S-matrix?}
In this letter we answer this question in the affirmative for the leading logarithmic soft theorem: 
we show for scalar QED and gravity that the $n=0$ soft theorem involving $S^{(\ln \omega)}_0$ in \eqref{logsoftexp} is the Ward identity of the subleading asymptotic symmetry whose conservation laws were derived in \cite{Campiglia:2019wxe,AtulBhatkar:2019vcb,Saha:2019tub} and which we prove to be generated by Noether charges that receive both classical and quantum long-range infrared corrections.

In the absence of long-range infrared effects, the tree-level soft factor $S_{0}$ would be the next term in the soft expansion beyond $S_{-1}$. 
Interestingly, while the structure of $S^{(\ln\omega)}_0$ is very different from $S_{0}$, there exists a heuristic relation \cite{Sahoo:2018lxl}
which hints that the logarithmic soft theorem may also be the Ward identity for the subleading asymptotic symmetry but with the associated charges receiving loop corrections. This is also suggested by the soft operator structure in the boost basis: while powers $\omega^n$ map to simple poles in $\Delta=-n$ for sub$^{n+1}$-leading tree-level soft theorems, logarithmic terms $\omega^n(\ln \omega)^{n+1}$ with $n\geq 0$ lead to poles of order $n+2$ at the {\it same} value $\Delta=-n$, and so both symmetries should be generated by $\mathcal O_{\Delta=-n}$ when loop corrections are accounted for \footnote{We thank E.~Trevisani for discussion on this point.}\footnote{Related recent work includes \cite{Krishna:2023ukw,Bhardwaj:2024wld}}.

Although there has been significant progress towards relating logarithmic soft theorems to asymptotic symmetries 
\cite{Campiglia:2019wxe,AtulBhatkar:2019vcb,Donnay:2022hkf,Agrawal:2023zea}, a first principles derivation from the covariant phase space perspective has so far remained elusive. A clear relationship between asymptotic symmetries and loop-exact universal soft theorems is required to establish that the infinite-dimensional soft algebras, such as $w_{1+{\infty}}$, are indeed the asymptotic symmetries of the (infrared-finite part of) the non-perturbative S-matrix.

Here we take an important step towards this goal: 
we construct regularized Noether charges
associated to the subleading asymptotic symmetries of massive scalar QED and perturbative gravity of the form
\begin{equation}
    Q^{\Lambda}=\ln \Lambda^{-1} \left(Q_H^{(\ln)}+Q_S^{(\ln)}\right)+\dots,
\end{equation} 
where $Q_H^{(\ln)}$ and $Q_S^{(\ln)}$ are the hard and soft charges computed on, respectively, the time-like and null boundary after accounting for long-range infrared effects, and show that, after stripping off the infrared regulator ($\Lambda \to 0$ for large $|\text{time}|\to \Lambda^{-1}$), their conservation between past and future asymptotic boundaries can be recognized as the leading logarithmic soft theorem.
This establishes from first principles the asymptotic symmetry interpretation of the universal all-loop exact logarithmic soft theorems. 

In this letter we present our main results, focusing on the case of scalar QED in section~\ref{ScalarQED} and summarizing our findings for perturbative gravity in section~\ref{Gravity}. We end with concluding remarks in section~\ref{Conclusion}.  A more detailed exposition of the asymptotic symmetries associated to logarithmic soft theorems will appear in longer companion papers~\cite{Choi:2024ajz,Choi:2024mac}.

\section{Scalar QED}
\label{ScalarQED}
We consider electromagnetic scattering of massive charged scalar particles. The asymptotic boundary for matter particles and gauge bosons is the disjoint union of time-like and null infinity, $i^{\pm}\, \cup\, {\cal I}^{\pm}$, with future ($+$) and past ($-$) components. 
In the covariant phase space formalism we compute the symplectic structure of the scalar field and the electromagnetic field whose fall-off at early and late times lead to long-range infrared effects which require a regularization procedure. Upon identifying a suitable (divergent) gauge parameter that smoothly interpolates across $i^{\pm}\, \cup\, {\cal I}^{\pm}$ we construct renormalized Noether charges whose conservation yields the leading logarithmic soft photon theorem.

\subsection{Subleading Asymptotic Symmetries}
\label{lslssq}

We analyze the generators of large $U(1)$ gauge transformations (LGT) at the future  asymptotic boundary $i^{+} \cup {\cal I}^{+}$ in Lorenz gauge; the analysis at the past boundary is exactly analogous. We start by identifying a suitable gauge parameter in the transformation of the gauge field ${\cal A}_\mu$ and the complex massive scalar field  $\phi$ minimally coupled to the gauge field,
\begin{equation}\label{deltaAphi}
\delta {\cal A}_\mu=\partial_\mu \epsilon^{\cal I^+},\quad \delta \phi=i e \epsilon^{i^+} \phi,
\end{equation}
that smoothly interpolates across $i^{+} \cup {\cal I}^{+}$.
In the future light-cone we use a hyperbolic slicing of Minkowski space parametrized by Euclidean AdS$_{3}$ coordinates $\tau=\sqrt{t^2-r^2}$ and $\rho=\frac{r}{\sqrt{t^2-r^2}}$ where $i^+$ is approached in the limit $\tau\to \infty$ at fixed $\rho$. The relationship between the momentum $\vec p$ of a massive particle and a point $(\rho, \hat{p})$ on $i^{+}$ is $\vec p/m=\rho\hat p$. 
Our ansatz for the large-$\tau$ behavior of the gauge parameter at $i^{+}$ is,
\begin{align}\label{lgtati+}
    \epsilon^{i^+}(\tau,y) \overset{\tau\to\infty}{=}\tau \bar\epsilon(y)+ O(\tfrac{\ln\tau}{\tau}),
\end{align}
where $y^\alpha=(\rho,\hat x)$ are the space-like coordinates on the three-dimensional hyperboloid $\cal H$ \mbox{(the ``blow up of $i^+$'')} and $\hat x \in S^2$. Terms of order $O(1)$ are omitted since these constitute the superphaserotations whose associated Ward identity is known to be Weinberg's soft photon theorem \cite{Campiglia:2015qka}.
In Lorenz gauge, $\bar\epsilon(y)$ solves \mbox{$( \triangle_{{\cal H}} - 3 ) \bar\epsilon = 0$},
where $\triangle_{\cal H}$ is the Laplacian on ${\cal H}$.
Using the bulk-to-boundary Green's function $G(y;\hat x)$ at $i^{+}$ \cite{Campiglia:2015qka,Campiglia:2015lxa}, we can parametrize $\bar\epsilon(y)$ in terms of a function $\epsilon(\hat x)$ on the $S^2$,
\begin{align}\label{greens12}
    \bar\epsilon(y)
    &=
        \int_{S^2} d^2\hat x\,
        G(y;\hat x)\epsilon(\hat x)
    .
\end{align}
In the large-$\rho$ limit we have the asymptotic behavior 
$\rho^{-1}G(\rho,\hat x;\hat x')\overset{\rho\to\infty}{=}\delta^{(2)}(\hat x-\hat x')$ 
so that using $\tau \rho=r$ we have
\begin{equation}
   \epsilon^{i^+}(\tau,\rho,\hat x)\overset{\rho\to\infty,\,\tau\to\infty}{=}r\epsilon(\hat x)+\dots.
\end{equation}

Massless particles reach future null infinity, \mbox{${\cal I}^{+}\simeq S^{2}\, \times\, {\mathbb R}$}, as $t+r\to \infty$ at $t-r=\text{fixed}$, and so we parametrize it by retarded Bondi coordinates $u=t-r$. 
At $\cal I^+$, the subleading symmetries in Lorenz gauge are generated by the following parameter \cite{Campiglia:2016hvg},
\begin{align}\label{lgtatscri}
    \epsilon^{\cal I^+}(u,r,\hat{x})
    &\overset{r\to\infty}=
    r\epsilon(\hat{x})
    + u (1 + \tfrac{1}{2}D^2) \epsilon(\hat{x})
    + O(\tfrac{\ln r}{r})
    ,
\end{align}
where $D^2$ is the Laplacian on the unit $S^2$.
Without loss of generality, we have chosen the same function $\epsilon(\hat x)$ on the $S^2$ so that at leading order $\epsilon^{i^+}=\epsilon^{\cal I^+}$.
Thus we have identified a large gauge parameter that smoothly extends across $i^{+} \cup {\cal I}^{+}$, with the one at $i^+$ being ``sourced'' by the one at $\cal I^+$.

  \subsection{Asymptotic Phase Space}
 \label{apssq}
A large gauge transformation with the parameter in~\eqref{lgtatscri} introduces to the angular component ${\cal A}_C$ of the gauge field, where $C$ is an index on the $S^2$, terms that are divergent at large $r$ and at large $u$ \cite{Peraza:2023ivy}.
It will be convenient to separate out these modes by the decomposition 
\begin{align}
\label{calAC}
{\cal A}_{C}(r,u,\hat x)\overset{r\to\infty}=\partial_{C}[r \alpha(\hat{x})+u \beta(\hat{x})]+A_{C}(u,\hat{x}),
\end{align}
where $\alpha$ and $\beta$ are Goldstone modes that transform homogeneously as \mbox{$\delta\alpha=\epsilon$} and \mbox{$\delta\beta=(1 +\frac{1}{2}D^2)\epsilon$} under subleading LGTs~\footnote{In the literature, LGTs with divergent (``overleading'') gauge parameters are sometimes referred to as overleading LGTs; here we instead we refer to them as subleading LGTs since they are related to subleading soft theorems.}, while
$A_{C}$ is the finite energy radiative data at ${\cal I}^{+}$ 
whose large-$u$ behavior is \cite{Campiglia:2019wxe}
\begin{align}\label{raddataatscri}
    A_C(u,\hat x) \overset{u\to\pm\infty}{=}
    A_C^{\pm(0)}(\hat x) + \frac{1}{|u|}A_C^{\pm(1)}(\hat x) + \dots,
\end{align}
where $A_C^{\pm(0)}$ is responsible for electromagnetic memory while $\frac{1}{|u|}A_C^{\pm(1)}$ gives the ``tail'' to this memory \cite{Saha:2019tub}. 
The asymptotic behavior of the gauge field at $i^+$ has a similar late-time behavior~\cite{Campiglia:2019wxe}
\begin{equation}\label{calAtau}
    {\cal A}_\tau(\tau,y)\overset{\tau\to\infty}{=}\frac1\tau \overset1A_\tau(y)+\dots .
\end{equation}
It is these infrared divergent terms at late time that are related to the virtual infrared divergence  in the Dyson \mbox{S-matrix}.
In~\cite{Campiglia:2015qka} it was shown that the fall-offs \eqref{calAtau} imply that the $\tau \to \infty$ asymptotics of the complex massive scalar field $\phi$ minimally coupled to the gauge field at $i^{+}$ has a phase $e^{ie\ln \tau \overset1A_\tau(y)}$ relative to the free field expression
and can be perturbatively expanded as~\cite{Campiglia:2015qka,Choi:2024mac}
\begin{align}\label{scalarati+}
    \phi(\tau, y)
    &=
	    \frac{\sqrt{m}}{2(2\pi)^{3/2}}
	    \sum_{n=0}^{\infty}
	    \bigg[
	    \frac{e^{-im\tau}}{\tau^{\frac32+n}}
	    \left(
	    	\overset \ln b{}_{n}(y)\ln\tau
	    	+ b_n(y)
	    \right)
	    \nonumber\\&
    	+ \frac{e^{im\tau}}{\tau^{\frac32+n}}
    	\left(
    		\overset \ln d{}^{\dagger}_{n}(y)\ln\tau
    		+ d^{\dagger}_{n}(y)
    	\right)
    	\bigg]\, +\, O(e^{2})
    .
\end{align}
The coefficients $\overset\ln b_n$ and $b_{n+1}$ ($\overset\ln d{}_n^\dagger$ and $d^\dagger_{n+1}$) for all $n\geq 0$ are fixed in terms of the lowest-order one $b\equiv b_0$ ($d^\dagger\equiv d^\dagger_0$) by the (interacting) equations of motion.

  \subsection{Logarithmic Hard and Soft Charges}
 \label{charges}
From the asymptotic behavior of the fields \eqref{calAC}
-\eqref{scalarati+}  we can compute the symplectic structure
\begin{equation}\label{Omegafuture}
    \Omega_{ i^{+}\cup {\cal I}^{+}}=\Omega_{i^{+}}+\Omega_{{\cal I}^{+}},
\end{equation}
where for the complex charged scalar field at $i^+$ we have
\begin{align}\label{sympati+}
    \Omega_{i^+}
    &=
        \lim_{\tau\rightarrow\infty}\int_{\cal H}d^3y\,
        \tau^3
        \left(
            - \delta\phi^\dagger \wedge \partial_\tau\delta\phi
            + \text{h.c.}
        \right),
\end{align}
while the radiative symplectic structure for the gauge field at  
${\cal I}^{+}$ is given by \cite{Peraza:2023ivy} 
\begin{align}\label{sympatscri+}
    \Omega_{\cal I^+}
    =
        \int_{{\cal I}^{+}} du\,d^2\hat x
        \,\gamma^{AB}
        \bigg(
        &\delta \partial_{u}A_A\wedge \delta A_B
        \nonumber\\&
        - \frac u2D^2 D_A\delta \partial_uA_B\wedge\delta\alpha
        \bigg),
\end{align}
with $\gamma_{AB}$ and $D_A$ denoting, respectively, the metric and the covariant derivative on the $S^2$. The first term in \eqref{sympatscri+} is the well-known symplectic structure for the finite-energy radiative data, while the second term can be traced back to the Goldstone mode in \eqref{calAC} induced by divergent LGTs.

The symplectic structure \eqref{Omegafuture} diverges at late times $\tau \to \infty$ and $u\to \infty$ and we regulate it by introducing a late-time cutoff $\Lambda^{-1}$.
The linear in $\tau\to \Lambda^{-1}$ divergence in \eqref{sympati+} can then be removed by adding a counterterm of the form 
\begin{align}\label{counterterm}
    \Omega^{\textrm{c.t.}}_{i^+}=-\Lambda^{-1}
    \int_{\cal H}d^3y\, \delta j^{(0)}_{\tau}(y)
    \wedge\delta\bar{\alpha}(y),
\end{align}
where $j_{\tau}^{(0)}(y)=-\frac{em^2}{2(2\pi)^3}[b^\dagger(y) b(y)-d^\dagger(y) d(y)]$ is the free field current and $\bar{\alpha}(y)=\int_{S^2} d^2\hat x\,G(y;\hat x)\alpha(\hat x)$. The renormalized symplectic structure at $i^+$ then diverges logarithmically as the cutoff is removed ($\Lambda \to 0$). Regulating the symplectic structure at $\cal I^+$ is more subtle: the $1/|u|$ tail in the radiative mode \eqref{raddataatscri} renders the integral over null infinity logarithmically divergent, and we regularize this by the same late-time cutoff $u\to \Lambda^{-1}$; the details will be explained in \cite{Choi:2024mac}. The bottom line is that the renormalization procedure for the symplectic structure in the union of time-like and null infinity \eqref{Omegafuture} can be recast as subtracting a ``corner term'' at $\cal I^+_-$ of the form
\begin{align}
    \Omega_{i^{+}\cup {\cal I}^{+}}^{\textrm{c.t.}}
    =
    -\Lambda^{-1}\int_{\cal I^+_-} d^2\hat{x}\,
    r^2\delta F_{ur}\wedge\delta\alpha
    .
\end{align}

Taking one variation in the (renormalized) symplectic structure on the future boundary to be the subleading LGTs \eqref{deltaAphi} parametrized by the gauge parameters~\eqref{lgtati+} and~\eqref{lgtatscri}, we can compute from $\Omega_{i^+\cup \cal I^+}(\delta,\delta_\epsilon)=\delta Q[\epsilon]$ the associated Noether charge $Q$. The charge can be expressed as the sum of hard and soft charges,
\begin{equation}
    Q=Q_H+Q_S,
\end{equation}
computed, respectively, from $\Omega_{i^+}$ and $\Omega_{\cal I^+}$.
The regularized total charge defined on $i^+\cup \cal I^+$ is then given by
\begin{align}\label{full_charge}
    Q^{\Lambda}[\epsilon]
    &=\nonumber
        \ln \Lambda^{-1}\, \left(
            Q^{(\ln)}_H[\bar\epsilon]
            + Q^{(\ln)}_S[\epsilon]
        \right)
        \nonumber\\&\quad
        + \left(Q^{(0)}_H[\bar\epsilon]
        + Q^{(0)}_S[\epsilon]\right)
        + \dots\,
    .
\end{align}
The subleading hard and soft charges in the second line are given by the finite expressions
\begin{align}
    Q^{(0)}_H[\bar\epsilon]
    &=
        -\frac{iem}{4(2\pi)^{3}} \int_{i^{+}} d^3y\,
        ({\cal D}^\alpha\bar\epsilon)
        \nonumber\\&\qquad\times
        \Big[
            b^\dagger \partial_\alpha b
            - (\partial_\alpha b^\dagger) b
            + (b\to d^\dagger)
        \Big]
    \nonumber,\\
    Q^{(0)}_S[\epsilon]
    &=
        -\frac12
        \int_{\cal I^+} du\, d^2\hat x \,
        \epsilon\,u\partial_u D^2D^C \cA_C
    ,
    \label{QS1}
\end{align}
where $\cA_C$ denotes the radiative data \eqref{raddataatscri} with the $1/u$ tail stripped off.
They are linear in the coupling $e$ and their conservation from past to future boundary corresponds to the tree-level subleading soft photon theorem (with factor $S_0$) of Low \cite{Campiglia:2019wxe}. The omitted terms $\dots$ in the second line of~\eqref{full_charge} contain loop-order corrections to these subleading charges (finite in the large-time limit $\Lambda\to0$) with the tree-level subleading charge recovered at linear order.

Our interest here focuses on the regularized charge $Q^{(\ln)}\equiv Q^{(\ln)}_H\, +\, Q^{(\ln)}_S$ in the first line of \eqref{full_charge} with
\begin{align}
Q^{(\ln)}_H[\bar\epsilon]
    &=
        \frac{e^{2}m}{2(2\pi)^{3}} \int_{i^{+}} d^{3}y  ( {\cal D}^{\alpha}\bar\epsilon) (\partial_{\alpha} \overset{1}{A}_{\tau} )
        (b^\dagger b + d^\dagger d)
    \nonumber,\\
    Q^{(\ln)}_S[\epsilon]
    &=
        -\frac12
        \int_{\cal I^+} du\,d^2\hat x \,
            \epsilon \,
            \partial_u (u^2\partial_u D^2D^C A_C)
    ,
    \label{QSlnw}
\end{align}
where $\cal D_\alpha$ is the covariant derivative on ${\cal H^+}$.
The same functional form is obtained for the counterpart of the charge on the past boundary for a gauge parameter which satisfies $\epsilon^{i^-}=\epsilon^{\cal I^-}$ at leading order (provided we choose again the same function on the sphere at $i^-$ and $\cal I^-$)
and so smoothly extends across $i^-\cup \cal I^-$.

\subsection{Loop Corrected Conservation Law}\label{LoopCharges}
The final step in our proof is to link the above charges to the logarithmic soft theorem. This is done by noticing 
that \eqref{QSlnw} and its counterpart on the past boundary 
are precisely the charges appearing in the conservation law between past $(-)$ and future $(+)$ boundary,
\begin{equation}
  	Q^{(\ln)}_{+} = Q^{(\ln)}_{-},
\end{equation}
which holds upon antipodal identification of the gauge parameters and fields between $\cal I^-_+$ and $\cal I^+_-$.
Upon quantization and for a judicious choice of function on the sphere, the authors of~\cite{Campiglia:2019wxe} showed that the Ward identity associated to $Q^{(\ln)}_\pm$ lands us precisely on the logarithmic soft photon theorem.  We emphasize that we have {\it derived} $Q^{(\ln)}_+$  from first principles as the Noether charge associated to the subleading asymptotic $U(1)$ symmetry. 

\section{Gravity}
\label{Gravity}

The asymptotics of gravitational radiative data coupled to massive sources in $D = 4$ dimensions has been an active subject of investigation for decades (see \cite{GIVPCD,tdamour1986,Kehrberger:2021uvf} and references therein).  A detailed derivation of the $D = 4$ superrotation flux at ${\cal I}$ in the non-linear theory in the Bondi gauge is under active investigation \cite{Geiller:2022vto,Geiller:2024amx,Geiller:2024ryw}. 
Here we consider the effective quantum theory of linear metric perturbations $h_{\mu\nu}$ around flat spacetime,
\begin{align}
    g_{\mu\nu} = \eta_{\mu\nu} + \kappa h_{\mu\nu}
    ,
    \qquad
    \kappa =\sqrt{32\pi G_N}
    ,
\end{align}
minimally coupled to a real massive scalar field~$\varphi$.
We work in harmonic (de Donder) gauge $\nabla^\nu h_{\mu\nu} - \frac12\nabla_\mu h=0$ where $h$ is the trace of the perturbation and $\nabla_\mu$ the spacetime covariant derivative.
The subleading soft graviton theorem related to superrotations receives loop corrections which manifest themselves as long-range infrared corrections in the superrotation charges
\footnote{As the gravitational soft theorems have been derived in harmonic gauge, we believe that the logarithmic divergence of the superrotation charge is a physical effect referred to as tail to the memory in \cite{Laddha:2018vbn}.}.
Here we summarize the main results of this analysis; the detailed derivation will appear in \cite{Choi:2024ajz}. Again we focus our discussion on the future boundary $i^+\cup \cal I^+$, while an analogous analysis follows for the past boundary $i^-\cup \cal I^-$.

Superrotations are parametrized by a vector field $Y^A(\hat x)$ at $\cal I^+$ \cite{Strominger:2017zoo} which can be smoothly extended to~$i^+$, where it satisfies  ${\cal D}_\alpha \bar Y^\alpha=0$, by smearing it with a vector Green's function \cite{Campiglia:2015lxa,Campiglia:2015kxa},
\begin{align}
    \bar Y^\alpha(y)
    &=
        \int_{S^2}d^2\hat x\,
        G^\alpha_A(y;\hat x)
        Y^A(\hat x)
    .
\end{align}
Superrotations on $i^+$ act as Lie derivative $\delta_Y\varphi = \bar Y^\alpha\partial_\alpha\varphi$.
A real massive scalar field $\varphi$ is expanded near $i^+$ analogous to \eqref{scalarati+} and the equations of motion $\nabla^2\varphi - m^2\varphi - \kappa h_{\mu\nu}\nabla^\mu\nabla^\nu \varphi=0$ fix all coefficients in terms of the lowest-order one $b\equiv b_0$.
The asymptotic behavior of the metric 
\begin{equation}
    h_{\tau\tau}(\tau,y)\overset{\tau\to\infty}{=}\frac1\tau \overset1h_{\tau\tau}(y)+\dots
\end{equation}
gives rise to logarithmic terms in the large-$\tau$ expansion of~$\varphi$.
The symplectic structure for the scalar field on $i^+$ is formally divergent, which we regulate with the large-time cutoff $\Lambda^{-1}$.
The gravitational radiative data at $\cal I^+$ is given by the shear $C_{AB}(u,\hat x)=\lim_{r\to\infty}\frac\kappa rh_{AB}(u,r,\hat x)$.
As $u\to\pm\infty$ the shear has the following behavior,
\begin{align}\label{shearatscri}
    C_{AB}(u,\hat x)\overset{u\to\pm\infty}{=}
    C_{AB}^{\pm(0)}(\hat x)
    + \frac1{|u|}C_{AB}^{\pm(1)}(\hat x)
    + \dots
    ,
\end{align}
where $C_{AB}^{\pm(0)}$ gives rise to gravitational displacement memory while $\frac1{|u|}C_{AB}^{\pm(1)}$ is its tail \cite{Laddha:2018vbn}. The $\frac1{|u|}$ term leads to a logarithmic divergence in the symplectic structure at $\cal I^+$ which we regulate with the same large-time cutoff $\Lambda^{-1}$.
The regularized %
superrotation charge is
\begin{equation}\label{Qgravity}
\badat{2}
     Q^{\Lambda}[Y]
    &=
        \ln \Lambda^{-1}\, \left(
            Q^{(\ln)}_H[\bar Y]
            + Q^{(\ln)}_S[Y]
        \right)
        \\&\quad
        + \left(Q^{(0)}_H[\bar Y]
        + Q^{(0)}_S[Y]\right)
        + \dots
    .
    \eadat
\end{equation}
The finite tree-level charges are given by \cite{Campiglia:2015kxa}
\begin{equation}\label{Q0gravity}
\badat{2}
    Q^{(0)}_H[\bar Y]
    &=
        \frac{im^2}{4(2\pi)^3}
        \int_{i^+}d^3y\,\bar Y^\alpha
        \left[
            b^\dagger \partial_\alpha b
            - (\partial_\alpha b^\dagger)b
        \right],\\
    Q^{(0)}_S[Y]
    &=
        -\frac{2}{\kappa^2}\int_{\cal I^+}du\,d^2\hat x\,
        D_z^3Y^z u\partial_u \cC^{zz}
        + \text{h.c.},
        \eadat
\end{equation}
whose associated conservation law between the past and future boundary corresponds upon quantization to the tree-level subleading soft graviton theorem~\cite{Campiglia:2015kxa}. Here $z=e^{i\phi}\tan\frac\theta2$ is the complex stereographic coordinate on the $S^2$, and $\cC^{zz}$ in \eqref{Q0gravity} denotes the shear \eqref{shearatscri} with the $1/u$ tail stripped off. The omitted terms $\dots$ in \eqref{Q0gravity} are finite in the $\Lambda\to0$ limit and of higher order in the coupling thus loop-correcting these subleading charges.
The regularized logarithmic charge $Q^{(\ln)}\equiv Q^{(\ln)}_H+Q^{(\ln)}_S$ is given by
\begin{equation}\label{Qlngravity}
\badat{2}
    Q^{(\ln)}_H[\bar Y]
    &=
        -\frac{\kappa m^3}{4(2\pi)^3}
        \int_{i^+}d^3y\, \bar Y^\alpha
            b^\dagger b \partial_\alpha \overset1h_{\tau\tau}
    ,\\
    Q^{(\ln)}_S[Y]
    &=
        -\frac{2}{\kappa^2}
        \int_{\cal I^+} du\, d^2\hat x\,
        D_z^3 Y^z
        \partial_u(u^2\partial_u C^{zz})
        + \text{h.c.}.
\eadat
\end{equation}
Upon antipodal identification of the fields and superrotation vector fields on  $\cal I^-_+$ and their counterparts on $\cal I^-_+$, the logarithmic superrotation charges satisfy $Q_+^{(\ln)}=Q_-^{(\ln)}$. 
This conservation law, upon judicious choice of superrotation vector field, matches the classical logarithmic soft graviton theorem \cite{Saha:2019tub} thus establishing its symmetry interpretation \footnote{There is an additional drag term that has to be treated separately, as we will explain in \cite{Choi:2024ajz}.}. The derivation of the quantized Ward identity is an open problem which we will address elsewhere \cite{elsewhere}.
In contrast to earlier literature \cite{Agrawal:2023zea}, our covariant phase space derivation achieves a clear split between the charges associated to the leading (Weinberg) and logarithmic (Sahoo-Sen) soft theorems, our regulator arises from the relevant infrared scale and can be removed in the end since the observables we computed, the charges $Q_\pm^{(\ln)}$ appearing in the conservation law, %
are finite. 

\section{Conclusion}\label{Conclusion}

In this letter we derived from first principles in a covariant phase space analysis the long-range infrared corrections to the subleading asymptotic symmetry charges in scalar QED and perturbative gravity. The regularized charges give rise to finite (regulator-independent) Ward identities which agree precisely with the leading logarithmic soft theorems of Sahoo-Sen.
We thereby established the symmetry interpretation of these universal all-loop exact soft theorems in gauge theory and gravity. 

Notice that the only difference between the expressions for the subleading tree-level soft charge in \eqref{QS1} and the logarithmic soft charge in \eqref{QSlnw} for scalar QED (or,  for  gravity, \eqref{Q0gravity} and \eqref{Qlngravity}) are the projectors, $u\partial_u$ and $\partial_u u^2\partial_u$, respectively. In the Ward identity, the two charges correspond to the same soft photon (graviton) insertion but the projectors pick out different terms in the soft expansion: $O(1)$ at tree-level and $O(\ln\omega)$ at loop level. This confirms the expectation raised in the introduction that the same operator is responsible for both symmetries. 
Moreover, we expect the procedure we have developed to apply to all universal logarithmic soft theorems. This raises the remarkable prospect of a tower of universal all-loop exact infinite-dimensional symmetries governing gauge theory and quantum gravity in asymptotically flat spacetimes.
We leave this study for future work.

\section*{Acknowledgments}
We would like to thank Sayali Bhatkar, Miguel Campiglia, Prahar Mitra, Biswajit Sahoo, Ashoke Sen, Adarsh S, Amit Suthar, Emilio Trevisani  and especially Marc Geiller and Javier Peraza for discussions.
SC and AP are supported by the European Research Council (ERC) under the European Union’s Horizon 2020 research and innovation programme (grant agreement No 852386). This work was supported by the Simons Collaboration on Celestial Holography.

\bibliographystyle{apsrev4-2}
\bibliography{references}

\begin{thebibliography}{80}%
\makeatletter
\providecommand \@ifxundefined [1]{%
 \@ifx{#1\undefined}
}%
\providecommand \@ifnum [1]{%
 \ifnum #1\expandafter \@firstoftwo
 \else \expandafter \@secondoftwo
 \fi
}%
\providecommand \@ifx [1]{%
 \ifx #1\expandafter \@firstoftwo
 \else \expandafter \@secondoftwo
 \fi
}%
\providecommand \natexlab [1]{#1}%
\providecommand \enquote  [1]{``#1''}%
\providecommand \bibnamefont  [1]{#1}%
\providecommand \bibfnamefont [1]{#1}%
\providecommand \citenamefont [1]{#1}%
\providecommand \href@noop [0]{\@secondoftwo}%
\providecommand \href [0]{\begingroup \@sanitize@url \@href}%
\providecommand \@href[1]{\@@startlink{#1}\@@href}%
\providecommand \@@href[1]{\endgroup#1\@@endlink}%
\providecommand \@sanitize@url [0]{\catcode `\\12\catcode `\$12\catcode
  `\&12\catcode `\#12\catcode `\^12\catcode `\_12\catcode `\%12\relax}%
\providecommand \@@startlink[1]{}%
\providecommand \@@endlink[0]{}%
\providecommand \url  [0]{\begingroup\@sanitize@url \@url }%
\providecommand \@url [1]{\endgroup\@href {#1}{\urlprefix }}%
\providecommand \urlprefix  [0]{URL }%
\providecommand \Eprint [0]{\href }%
\providecommand \doibase [0]{https://doi.org/}%
\providecommand \selectlanguage [0]{\@gobble}%
\providecommand \bibinfo  [0]{\@secondoftwo}%
\providecommand \bibfield  [0]{\@secondoftwo}%
\providecommand \translation [1]{[#1]}%
\providecommand \BibitemOpen [0]{}%
\providecommand \bibitemStop [0]{}%
\providecommand \bibitemNoStop [0]{.\EOS\space}%
\providecommand \EOS [0]{\spacefactor3000\relax}%
\providecommand \BibitemShut  [1]{\csname bibitem#1\endcsname}%
\let\auto@bib@innerbib\@empty
\bibitem [{\citenamefont {Low}(1958)}]{Low:1958sn}%
  \BibitemOpen
  \bibfield  {author} {\bibinfo {author} {\bibfnamefont {F.~E.}\ \bibnamefont
  {Low}},\ }\bibfield  {title} {\emph {\bibinfo {title} {{Bremsstrahlung of
  very low-energy quanta in elementary particle collisions}}},\ }\href
  {https://doi.org/10.1103/PhysRev.110.974} {\bibfield  {journal} {\bibinfo
  {journal} {Phys. Rev.}\ }\textbf {\bibinfo {volume} {110}},\ \bibinfo {pages}
  {974} (\bibinfo {year} {1958})}\BibitemShut {NoStop}%
\bibitem [{\citenamefont {Weinberg}(1965)}]{Weinberg:1965nx}%
  \BibitemOpen
  \bibfield  {author} {\bibinfo {author} {\bibfnamefont {S.}~\bibnamefont
  {Weinberg}},\ }\bibfield  {title} {\emph {\bibinfo {title} {{Infrared photons
  and gravitons}}},\ }\href {https://doi.org/10.1103/PhysRev.140.B516}
  {\bibfield  {journal} {\bibinfo  {journal} {Phys. Rev.}\ }\textbf {\bibinfo
  {volume} {140}},\ \bibinfo {pages} {B516} (\bibinfo {year}
  {1965})}\BibitemShut {NoStop}%
\bibitem [{\citenamefont {Cachazo}\ and\ \citenamefont
  {Strominger}(2014)}]{Cachazo:2014fwa}%
  \BibitemOpen
  \bibfield  {author} {\bibinfo {author} {\bibfnamefont {F.}~\bibnamefont
  {Cachazo}}\ and\ \bibinfo {author} {\bibfnamefont {A.}~\bibnamefont
  {Strominger}},\ }\bibfield  {title} {\emph {\bibinfo {title} {{Evidence for a
  New Soft Graviton Theorem}}},\ }\href@noop {} {\  (\bibinfo {year} {2014})},\
  \Eprint {https://arxiv.org/abs/1404.4091} {arXiv:1404.4091 [hep-th]}
  \BibitemShut {NoStop}%
\bibitem [{\citenamefont {Hamada}\ and\ \citenamefont
  {Shiu}(2018)}]{Hamada:2018vrw}%
  \BibitemOpen
  \bibfield  {author} {\bibinfo {author} {\bibfnamefont {Y.}~\bibnamefont
  {Hamada}}\ and\ \bibinfo {author} {\bibfnamefont {G.}~\bibnamefont {Shiu}},\
  }\bibfield  {title} {\emph {\bibinfo {title} {{Infinite Set of Soft Theorems
  in Gauge-Gravity Theories as Ward-Takahashi Identities}}},\ }\href
  {https://doi.org/10.1103/PhysRevLett.120.201601} {\bibfield  {journal}
  {\bibinfo  {journal} {Phys. Rev. Lett.}\ }\textbf {\bibinfo {volume} {120}},\
  \bibinfo {pages} {201601} (\bibinfo {year} {2018})},\ \Eprint
  {https://arxiv.org/abs/1801.05528} {arXiv:1801.05528 [hep-th]} \BibitemShut
  {NoStop}%
\bibitem [{\citenamefont {Li}\ \emph {et~al.}(2018)\citenamefont {Li},
  \citenamefont {Lin},\ and\ \citenamefont {Zhang}}]{Li:2018gnc}%
  \BibitemOpen
  \bibfield  {author} {\bibinfo {author} {\bibfnamefont {Z.-Z.}\ \bibnamefont
  {Li}}, \bibinfo {author} {\bibfnamefont {H.-H.}\ \bibnamefont {Lin}},\ and\
  \bibinfo {author} {\bibfnamefont {S.-Q.}\ \bibnamefont {Zhang}},\ }\bibfield
  {title} {\emph {\bibinfo {title} {{Infinite Soft Theorems from Gauge
  Symmetry}}},\ }\href {https://doi.org/10.1103/PhysRevD.98.045004} {\bibfield
  {journal} {\bibinfo  {journal} {Phys. Rev. D}\ }\textbf {\bibinfo {volume}
  {98}},\ \bibinfo {pages} {045004} (\bibinfo {year} {2018})},\ \Eprint
  {https://arxiv.org/abs/1802.03148} {arXiv:1802.03148 [hep-th]} \BibitemShut
  {NoStop}%
\bibitem [{\citenamefont {Elvang}\ \emph {et~al.}(2017)\citenamefont {Elvang},
  \citenamefont {Jones},\ and\ \citenamefont {Naculich}}]{Elvang:2016qvq}%
  \BibitemOpen
  \bibfield  {author} {\bibinfo {author} {\bibfnamefont {H.}~\bibnamefont
  {Elvang}}, \bibinfo {author} {\bibfnamefont {C.~R.~T.}\ \bibnamefont
  {Jones}},\ and\ \bibinfo {author} {\bibfnamefont {S.~G.}\ \bibnamefont
  {Naculich}},\ }\bibfield  {title} {\emph {\bibinfo {title} {{Soft Photon and
  Graviton Theorems in Effective Field Theory}}},\ }\href
  {https://doi.org/10.1103/PhysRevLett.118.231601} {\bibfield  {journal}
  {\bibinfo  {journal} {Phys. Rev. Lett.}\ }\textbf {\bibinfo {volume} {118}},\
  \bibinfo {pages} {231601} (\bibinfo {year} {2017})},\ \Eprint
  {https://arxiv.org/abs/1611.07534} {arXiv:1611.07534 [hep-th]} \BibitemShut
  {NoStop}%
\bibitem [{\citenamefont {Ghosh}\ and\ \citenamefont
  {Sahoo}(2022)}]{Ghosh:2021bam}%
  \BibitemOpen
  \bibfield  {author} {\bibinfo {author} {\bibfnamefont {D.}~\bibnamefont
  {Ghosh}}\ and\ \bibinfo {author} {\bibfnamefont {B.}~\bibnamefont {Sahoo}},\
  }\bibfield  {title} {\emph {\bibinfo {title} {{Spin-dependent gravitational
  tail memory in $D=4$}}},\ }\href
  {https://doi.org/10.1103/PhysRevD.105.025024} {\bibfield  {journal} {\bibinfo
   {journal} {Phys. Rev. D}\ }\textbf {\bibinfo {volume} {105}},\ \bibinfo
  {pages} {025024} (\bibinfo {year} {2022})},\ \Eprint
  {https://arxiv.org/abs/2106.10741} {arXiv:2106.10741 [hep-th]} \BibitemShut
  {NoStop}%
\bibitem [{\citenamefont {Laddha}\ and\ \citenamefont
  {Sen}(2017)}]{Laddha:2017ygw}%
  \BibitemOpen
  \bibfield  {author} {\bibinfo {author} {\bibfnamefont {A.}~\bibnamefont
  {Laddha}}\ and\ \bibinfo {author} {\bibfnamefont {A.}~\bibnamefont {Sen}},\
  }\bibfield  {title} {\emph {\bibinfo {title} {{Sub-subleading Soft Graviton
  Theorem in Generic Theories of Quantum Gravity}}},\ }\href
  {https://doi.org/10.1007/JHEP10(2017)065} {\bibfield  {journal} {\bibinfo
  {journal} {JHEP}\ }\textbf {\bibinfo {volume} {10}},\ \bibinfo {pages}
  {065}},\ \Eprint {https://arxiv.org/abs/1706.00759} {arXiv:1706.00759
  [hep-th]} \BibitemShut {NoStop}%
\bibitem [{\citenamefont {Krishna}\ and\ \citenamefont
  {Sahoo}(2023)}]{Krishna:2023fxg}%
  \BibitemOpen
  \bibfield  {author} {\bibinfo {author} {\bibfnamefont {H.}~\bibnamefont
  {Krishna}}\ and\ \bibinfo {author} {\bibfnamefont {B.}~\bibnamefont
  {Sahoo}},\ }\bibfield  {title} {\emph {\bibinfo {title} {{Universality of
  loop corrected soft theorems in 4d}}},\ }\href
  {https://doi.org/10.1007/JHEP11(2023)233} {\bibfield  {journal} {\bibinfo
  {journal} {JHEP}\ }\textbf {\bibinfo {volume} {11}},\ \bibinfo {pages}
  {233}},\ \Eprint {https://arxiv.org/abs/2308.16807} {arXiv:2308.16807
  [hep-th]} \BibitemShut {NoStop}%
\bibitem [{\citenamefont {Strominger}(2014)}]{Strominger:2013jfa}%
  \BibitemOpen
  \bibfield  {author} {\bibinfo {author} {\bibfnamefont {A.}~\bibnamefont
  {Strominger}},\ }\bibfield  {title} {\emph {\bibinfo {title} {{On BMS
  Invariance of Gravitational Scattering}}},\ }\href
  {https://doi.org/10.1007/JHEP07(2014)152} {\bibfield  {journal} {\bibinfo
  {journal} {JHEP}\ }\textbf {\bibinfo {volume} {07}},\ \bibinfo {pages}
  {152}},\ \Eprint {https://arxiv.org/abs/1312.2229} {arXiv:1312.2229 [hep-th]}
  \BibitemShut {NoStop}%
\bibitem [{\citenamefont {He}\ \emph {et~al.}(2015)\citenamefont {He},
  \citenamefont {Lysov}, \citenamefont {Mitra},\ and\ \citenamefont
  {Strominger}}]{He:2014laa}%
  \BibitemOpen
  \bibfield  {author} {\bibinfo {author} {\bibfnamefont {T.}~\bibnamefont
  {He}}, \bibinfo {author} {\bibfnamefont {V.}~\bibnamefont {Lysov}}, \bibinfo
  {author} {\bibfnamefont {P.}~\bibnamefont {Mitra}},\ and\ \bibinfo {author}
  {\bibfnamefont {A.}~\bibnamefont {Strominger}},\ }\bibfield  {title} {\emph
  {\bibinfo {title} {{BMS supertranslations and Weinberg\textquoteright{}s soft
  graviton theorem}}},\ }\href {https://doi.org/10.1007/JHEP05(2015)151}
  {\bibfield  {journal} {\bibinfo  {journal} {JHEP}\ }\textbf {\bibinfo
  {volume} {05}},\ \bibinfo {pages} {151}},\ \Eprint
  {https://arxiv.org/abs/1401.7026} {arXiv:1401.7026 [hep-th]} \BibitemShut
  {NoStop}%
\bibitem [{\citenamefont {Kapec}\ \emph {et~al.}(2014)\citenamefont {Kapec},
  \citenamefont {Lysov}, \citenamefont {Pasterski},\ and\ \citenamefont
  {Strominger}}]{Kapec:2014opa}%
  \BibitemOpen
  \bibfield  {author} {\bibinfo {author} {\bibfnamefont {D.}~\bibnamefont
  {Kapec}}, \bibinfo {author} {\bibfnamefont {V.}~\bibnamefont {Lysov}},
  \bibinfo {author} {\bibfnamefont {S.}~\bibnamefont {Pasterski}},\ and\
  \bibinfo {author} {\bibfnamefont {A.}~\bibnamefont {Strominger}},\ }\bibfield
   {title} {\emph {\bibinfo {title} {{Semiclassical Virasoro symmetry of the
  quantum gravity $ \mathcal{S}$-matrix}}},\ }\href
  {https://doi.org/10.1007/JHEP08(2014)058} {\bibfield  {journal} {\bibinfo
  {journal} {JHEP}\ }\textbf {\bibinfo {volume} {08}},\ \bibinfo {pages}
  {058}},\ \Eprint {https://arxiv.org/abs/1406.3312} {arXiv:1406.3312 [hep-th]}
  \BibitemShut {NoStop}%
\bibitem [{\citenamefont {Lysov}\ \emph {et~al.}(2014)\citenamefont {Lysov},
  \citenamefont {Pasterski},\ and\ \citenamefont {Strominger}}]{Lysov:2014csa}%
  \BibitemOpen
  \bibfield  {author} {\bibinfo {author} {\bibfnamefont {V.}~\bibnamefont
  {Lysov}}, \bibinfo {author} {\bibfnamefont {S.}~\bibnamefont {Pasterski}},\
  and\ \bibinfo {author} {\bibfnamefont {A.}~\bibnamefont {Strominger}},\
  }\bibfield  {title} {\emph {\bibinfo {title} {{Low\textquoteright{}s
  Subleading Soft Theorem as a Symmetry of QED}}},\ }\href
  {https://doi.org/10.1103/PhysRevLett.113.111601} {\bibfield  {journal}
  {\bibinfo  {journal} {Phys. Rev. Lett.}\ }\textbf {\bibinfo {volume} {113}},\
  \bibinfo {pages} {111601} (\bibinfo {year} {2014})},\ \Eprint
  {https://arxiv.org/abs/1407.3814} {arXiv:1407.3814 [hep-th]} \BibitemShut
  {NoStop}%
\bibitem [{\citenamefont {He}\ \emph {et~al.}(2014)\citenamefont {He},
  \citenamefont {Mitra}, \citenamefont {Porfyriadis},\ and\ \citenamefont
  {Strominger}}]{He:2014cra}%
  \BibitemOpen
  \bibfield  {author} {\bibinfo {author} {\bibfnamefont {T.}~\bibnamefont
  {He}}, \bibinfo {author} {\bibfnamefont {P.}~\bibnamefont {Mitra}}, \bibinfo
  {author} {\bibfnamefont {A.~P.}\ \bibnamefont {Porfyriadis}},\ and\ \bibinfo
  {author} {\bibfnamefont {A.}~\bibnamefont {Strominger}},\ }\bibfield  {title}
  {\emph {\bibinfo {title} {{New Symmetries of Massless QED}}},\ }\href
  {https://doi.org/10.1007/JHEP10(2014)112} {\bibfield  {journal} {\bibinfo
  {journal} {JHEP}\ }\textbf {\bibinfo {volume} {10}},\ \bibinfo {pages}
  {112}},\ \Eprint {https://arxiv.org/abs/1407.3789} {arXiv:1407.3789 [hep-th]}
  \BibitemShut {NoStop}%
\bibitem [{\citenamefont {Kapec}\ \emph
  {et~al.}(2017{\natexlab{a}})\citenamefont {Kapec}, \citenamefont {Pate},\
  and\ \citenamefont {Strominger}}]{Kapec:2015ena}%
  \BibitemOpen
  \bibfield  {author} {\bibinfo {author} {\bibfnamefont {D.}~\bibnamefont
  {Kapec}}, \bibinfo {author} {\bibfnamefont {M.}~\bibnamefont {Pate}},\ and\
  \bibinfo {author} {\bibfnamefont {A.}~\bibnamefont {Strominger}},\ }\bibfield
   {title} {\emph {\bibinfo {title} {{New Symmetries of QED}}},\ }\href
  {https://doi.org/10.4310/ATMP.2017.v21.n7.a7} {\bibfield  {journal} {\bibinfo
   {journal} {Adv. Theor. Math. Phys.}\ }\textbf {\bibinfo {volume} {21}},\
  \bibinfo {pages} {1769} (\bibinfo {year} {2017}{\natexlab{a}})},\ \Eprint
  {https://arxiv.org/abs/1506.02906} {arXiv:1506.02906 [hep-th]} \BibitemShut
  {NoStop}%
\bibitem [{\citenamefont {Campiglia}\ and\ \citenamefont
  {Laddha}(2015{\natexlab{a}})}]{Campiglia:2015qka}%
  \BibitemOpen
  \bibfield  {author} {\bibinfo {author} {\bibfnamefont {M.}~\bibnamefont
  {Campiglia}}\ and\ \bibinfo {author} {\bibfnamefont {A.}~\bibnamefont
  {Laddha}},\ }\bibfield  {title} {\emph {\bibinfo {title} {{Asymptotic
  symmetries of QED and Weinberg\textquoteright{}s soft photon theorem}}},\
  }\href {https://doi.org/10.1007/JHEP07(2015)115} {\bibfield  {journal}
  {\bibinfo  {journal} {JHEP}\ }\textbf {\bibinfo {volume} {07}},\ \bibinfo
  {pages} {115}},\ \Eprint {https://arxiv.org/abs/1505.05346} {arXiv:1505.05346
  [hep-th]} \BibitemShut {NoStop}%
\bibitem [{\citenamefont {Campiglia}\ and\ \citenamefont
  {Laddha}(2016)}]{Campiglia:2016hvg}%
  \BibitemOpen
  \bibfield  {author} {\bibinfo {author} {\bibfnamefont {M.}~\bibnamefont
  {Campiglia}}\ and\ \bibinfo {author} {\bibfnamefont {A.}~\bibnamefont
  {Laddha}},\ }\bibfield  {title} {\emph {\bibinfo {title} {{Subleading soft
  photons and large gauge transformations}}},\ }\href
  {https://doi.org/10.1007/JHEP11(2016)012} {\bibfield  {journal} {\bibinfo
  {journal} {JHEP}\ }\textbf {\bibinfo {volume} {11}},\ \bibinfo {pages}
  {012}},\ \Eprint {https://arxiv.org/abs/1605.09677} {arXiv:1605.09677
  [hep-th]} \BibitemShut {NoStop}%
\bibitem [{\citenamefont {Campiglia}\ and\ \citenamefont
  {Laddha}(2015{\natexlab{b}})}]{Campiglia:2015kxa}%
  \BibitemOpen
  \bibfield  {author} {\bibinfo {author} {\bibfnamefont {M.}~\bibnamefont
  {Campiglia}}\ and\ \bibinfo {author} {\bibfnamefont {A.}~\bibnamefont
  {Laddha}},\ }\bibfield  {title} {\emph {\bibinfo {title} {{Asymptotic
  symmetries of gravity and soft theorems for massive particles}}},\ }\href
  {https://doi.org/10.1007/JHEP12(2015)094} {\bibfield  {journal} {\bibinfo
  {journal} {JHEP}\ }\textbf {\bibinfo {volume} {12}},\ \bibinfo {pages}
  {094}},\ \Eprint {https://arxiv.org/abs/1509.01406} {arXiv:1509.01406
  [hep-th]} \BibitemShut {NoStop}%
\bibitem [{\citenamefont {Campiglia}\ and\ \citenamefont
  {Laddha}(2014)}]{Campiglia:2014yka}%
  \BibitemOpen
  \bibfield  {author} {\bibinfo {author} {\bibfnamefont {M.}~\bibnamefont
  {Campiglia}}\ and\ \bibinfo {author} {\bibfnamefont {A.}~\bibnamefont
  {Laddha}},\ }\bibfield  {title} {\emph {\bibinfo {title} {{Asymptotic
  symmetries and subleading soft graviton theorem}}},\ }\href
  {https://doi.org/10.1103/PhysRevD.90.124028} {\bibfield  {journal} {\bibinfo
  {journal} {Phys. Rev. D}\ }\textbf {\bibinfo {volume} {90}},\ \bibinfo
  {pages} {124028} (\bibinfo {year} {2014})},\ \Eprint
  {https://arxiv.org/abs/1408.2228} {arXiv:1408.2228 [hep-th]} \BibitemShut
  {NoStop}%
\bibitem [{Note1()}]{Note1}%
  \BibitemOpen
  \bibinfo {note} {Asymptotic symmetries in gauge theory are generically called
  large gauge transformations. In analogy to gravity, we refer to the leading
  one as superphaserotation.}\BibitemShut {Stop}%
\bibitem [{\citenamefont {Campiglia}\ and\ \citenamefont
  {Laddha}(2019{\natexlab{a}})}]{Campiglia:2018dyi}%
  \BibitemOpen
  \bibfield  {author} {\bibinfo {author} {\bibfnamefont {M.}~\bibnamefont
  {Campiglia}}\ and\ \bibinfo {author} {\bibfnamefont {A.}~\bibnamefont
  {Laddha}},\ }\bibfield  {title} {\emph {\bibinfo {title} {{Asymptotic charges
  in massless QED revisited: A view from Spatial Infinity}}},\ }\href
  {https://doi.org/10.1007/JHEP05(2019)207} {\bibfield  {journal} {\bibinfo
  {journal} {JHEP}\ }\textbf {\bibinfo {volume} {05}},\ \bibinfo {pages}
  {207}},\ \Eprint {https://arxiv.org/abs/1810.04619} {arXiv:1810.04619
  [hep-th]} \BibitemShut {NoStop}%
\bibitem [{\citenamefont {Freidel}\ \emph {et~al.}(2022)\citenamefont
  {Freidel}, \citenamefont {Pranzetti},\ and\ \citenamefont
  {Raclariu}}]{Freidel:2021ytz}%
  \BibitemOpen
  \bibfield  {author} {\bibinfo {author} {\bibfnamefont {L.}~\bibnamefont
  {Freidel}}, \bibinfo {author} {\bibfnamefont {D.}~\bibnamefont {Pranzetti}},\
  and\ \bibinfo {author} {\bibfnamefont {A.-M.}\ \bibnamefont {Raclariu}},\
  }\bibfield  {title} {\emph {\bibinfo {title} {{Higher spin dynamics in
  gravity and w1+\ensuremath{\infty} celestial symmetries}}},\ }\href
  {https://doi.org/10.1103/PhysRevD.106.086013} {\bibfield  {journal} {\bibinfo
   {journal} {Phys. Rev. D}\ }\textbf {\bibinfo {volume} {106}},\ \bibinfo
  {pages} {086013} (\bibinfo {year} {2022})},\ \Eprint
  {https://arxiv.org/abs/2112.15573} {arXiv:2112.15573 [hep-th]} \BibitemShut
  {NoStop}%
\bibitem [{\citenamefont {Raclariu}(2021)}]{Raclariu:2021zjz}%
  \BibitemOpen
  \bibfield  {author} {\bibinfo {author} {\bibfnamefont {A.-M.}\ \bibnamefont
  {Raclariu}},\ }\bibfield  {title} {\emph {\bibinfo {title} {{Lectures on
  Celestial Holography}}},\ }\href@noop {} {\  (\bibinfo {year} {2021})},\
  \Eprint {https://arxiv.org/abs/2107.02075} {arXiv:2107.02075 [hep-th]}
  \BibitemShut {NoStop}%
\bibitem [{\citenamefont {Prema}\ \emph {et~al.}(2022)\citenamefont {Prema},
  \citenamefont {Comp\`ere}, \citenamefont {Pipolo~de Gioia}, \citenamefont
  {Mol},\ and\ \citenamefont {Swidler}}]{Prema:2021sjp}%
  \BibitemOpen
  \bibfield  {author} {\bibinfo {author} {\bibfnamefont {A.~B.}\ \bibnamefont
  {Prema}}, \bibinfo {author} {\bibfnamefont {G.}~\bibnamefont {Comp\`ere}},
  \bibinfo {author} {\bibfnamefont {L.}~\bibnamefont {Pipolo~de Gioia}},
  \bibinfo {author} {\bibfnamefont {I.}~\bibnamefont {Mol}},\ and\ \bibinfo
  {author} {\bibfnamefont {B.}~\bibnamefont {Swidler}},\ }\bibfield  {title}
  {\emph {\bibinfo {title} {{Celestial holography: Lectures on asymptotic
  symmetries}}},\ }\href {https://doi.org/10.21468/SciPostPhysLectNotes.47}
  {\bibfield  {journal} {\bibinfo  {journal} {SciPost Phys. Lect. Notes}\
  }\textbf {\bibinfo {volume} {47}},\ \bibinfo {pages} {1} (\bibinfo {year}
  {2022})},\ \Eprint {https://arxiv.org/abs/2109.00997} {arXiv:2109.00997
  [hep-th]} \BibitemShut {NoStop}%
\bibitem [{\citenamefont {Pasterski}(2021)}]{Pasterski:2021rjz}%
  \BibitemOpen
  \bibfield  {author} {\bibinfo {author} {\bibfnamefont {S.}~\bibnamefont
  {Pasterski}},\ }\bibfield  {title} {\emph {\bibinfo {title} {{Lectures on
  celestial amplitudes}}},\ }\href
  {https://doi.org/10.1140/epjc/s10052-021-09846-7} {\bibfield  {journal}
  {\bibinfo  {journal} {Eur. Phys. J. C}\ }\textbf {\bibinfo {volume} {81}},\
  \bibinfo {pages} {1062} (\bibinfo {year} {2021})},\ \Eprint
  {https://arxiv.org/abs/2108.04801} {arXiv:2108.04801 [hep-th]} \BibitemShut
  {NoStop}%
\bibitem [{\citenamefont {McLoughlin}\ \emph {et~al.}(2022)\citenamefont
  {McLoughlin}, \citenamefont {Puhm},\ and\ \citenamefont
  {Raclariu}}]{McLoughlin:2022ljp}%
  \BibitemOpen
  \bibfield  {author} {\bibinfo {author} {\bibfnamefont {T.}~\bibnamefont
  {McLoughlin}}, \bibinfo {author} {\bibfnamefont {A.}~\bibnamefont {Puhm}},\
  and\ \bibinfo {author} {\bibfnamefont {A.-M.}\ \bibnamefont {Raclariu}},\
  }\bibfield  {title} {\emph {\bibinfo {title} {{The SAGEX review on scattering
  amplitudes chapter 11: soft theorems and celestial amplitudes}}},\ }\href
  {https://doi.org/10.1088/1751-8121/ac9a40} {\bibfield  {journal} {\bibinfo
  {journal} {J. Phys. A}\ }\textbf {\bibinfo {volume} {55}},\ \bibinfo {pages}
  {443012} (\bibinfo {year} {2022})},\ \Eprint
  {https://arxiv.org/abs/2203.13022} {arXiv:2203.13022 [hep-th]} \BibitemShut
  {NoStop}%
\bibitem [{\citenamefont {Donnay}(2023)}]{Donnay:2023mrd}%
  \BibitemOpen
  \bibfield  {author} {\bibinfo {author} {\bibfnamefont {L.}~\bibnamefont
  {Donnay}},\ }\bibfield  {title} {\emph {\bibinfo {title} {{Celestial
  holography: An asymptotic symmetry perspective}}},\ }\href@noop {} {\
  (\bibinfo {year} {2023})},\ \Eprint {https://arxiv.org/abs/2310.12922}
  {arXiv:2310.12922 [hep-th]} \BibitemShut {NoStop}%
\bibitem [{\citenamefont {de~Boer}\ and\ \citenamefont
  {Solodukhin}(2003)}]{deBoer:2003vf}%
  \BibitemOpen
  \bibfield  {author} {\bibinfo {author} {\bibfnamefont {J.}~\bibnamefont
  {de~Boer}}\ and\ \bibinfo {author} {\bibfnamefont {S.~N.}\ \bibnamefont
  {Solodukhin}},\ }\bibfield  {title} {\emph {\bibinfo {title} {{A Holographic
  reduction of Minkowski space-time}}},\ }\href
  {https://doi.org/10.1016/S0550-3213(03)00494-2} {\bibfield  {journal}
  {\bibinfo  {journal} {Nucl. Phys. B}\ }\textbf {\bibinfo {volume} {665}},\
  \bibinfo {pages} {545} (\bibinfo {year} {2003})},\ \Eprint
  {https://arxiv.org/abs/hep-th/0303006} {arXiv:hep-th/0303006} \BibitemShut
  {NoStop}%
\bibitem [{\citenamefont {Cheung}\ \emph {et~al.}(2017)\citenamefont {Cheung},
  \citenamefont {de~la Fuente},\ and\ \citenamefont
  {Sundrum}}]{Cheung:2016iub}%
  \BibitemOpen
  \bibfield  {author} {\bibinfo {author} {\bibfnamefont {C.}~\bibnamefont
  {Cheung}}, \bibinfo {author} {\bibfnamefont {A.}~\bibnamefont {de~la
  Fuente}},\ and\ \bibinfo {author} {\bibfnamefont {R.}~\bibnamefont
  {Sundrum}},\ }\bibfield  {title} {\emph {\bibinfo {title} {{4D scattering
  amplitudes and asymptotic symmetries from 2D CFT}}},\ }\href
  {https://doi.org/10.1007/JHEP01(2017)112} {\bibfield  {journal} {\bibinfo
  {journal} {JHEP}\ }\textbf {\bibinfo {volume} {01}},\ \bibinfo {pages}
  {112}},\ \Eprint {https://arxiv.org/abs/1609.00732} {arXiv:1609.00732
  [hep-th]} \BibitemShut {NoStop}%
\bibitem [{\citenamefont {Pasterski}\ \emph
  {et~al.}(2017{\natexlab{a}})\citenamefont {Pasterski}, \citenamefont {Shao},\
  and\ \citenamefont {Strominger}}]{Pasterski:2016qvg}%
  \BibitemOpen
  \bibfield  {author} {\bibinfo {author} {\bibfnamefont {S.}~\bibnamefont
  {Pasterski}}, \bibinfo {author} {\bibfnamefont {S.-H.}\ \bibnamefont
  {Shao}},\ and\ \bibinfo {author} {\bibfnamefont {A.}~\bibnamefont
  {Strominger}},\ }\bibfield  {title} {\emph {\bibinfo {title} {{Flat Space
  Amplitudes and Conformal Symmetry of the Celestial Sphere}}},\ }\href
  {https://doi.org/10.1103/PhysRevD.96.065026} {\bibfield  {journal} {\bibinfo
  {journal} {Phys. Rev. D}\ }\textbf {\bibinfo {volume} {96}},\ \bibinfo
  {pages} {065026} (\bibinfo {year} {2017}{\natexlab{a}})},\ \Eprint
  {https://arxiv.org/abs/1701.00049} {arXiv:1701.00049 [hep-th]} \BibitemShut
  {NoStop}%
\bibitem [{\citenamefont {Pasterski}\ and\ \citenamefont
  {Shao}(2017)}]{Pasterski:2017kqt}%
  \BibitemOpen
  \bibfield  {author} {\bibinfo {author} {\bibfnamefont {S.}~\bibnamefont
  {Pasterski}}\ and\ \bibinfo {author} {\bibfnamefont {S.-H.}\ \bibnamefont
  {Shao}},\ }\bibfield  {title} {\emph {\bibinfo {title} {{Conformal basis for
  flat space amplitudes}}},\ }\href
  {https://doi.org/10.1103/PhysRevD.96.065022} {\bibfield  {journal} {\bibinfo
  {journal} {Phys. Rev. D}\ }\textbf {\bibinfo {volume} {96}},\ \bibinfo
  {pages} {065022} (\bibinfo {year} {2017})},\ \Eprint
  {https://arxiv.org/abs/1705.01027} {arXiv:1705.01027 [hep-th]} \BibitemShut
  {NoStop}%
\bibitem [{\citenamefont {Pasterski}\ \emph
  {et~al.}(2017{\natexlab{b}})\citenamefont {Pasterski}, \citenamefont {Shao},\
  and\ \citenamefont {Strominger}}]{Pasterski:2017ylz}%
  \BibitemOpen
  \bibfield  {author} {\bibinfo {author} {\bibfnamefont {S.}~\bibnamefont
  {Pasterski}}, \bibinfo {author} {\bibfnamefont {S.-H.}\ \bibnamefont
  {Shao}},\ and\ \bibinfo {author} {\bibfnamefont {A.}~\bibnamefont
  {Strominger}},\ }\bibfield  {title} {\emph {\bibinfo {title} {{Gluon
  Amplitudes as 2d Conformal Correlators}}},\ }\href
  {https://doi.org/10.1103/PhysRevD.96.085006} {\bibfield  {journal} {\bibinfo
  {journal} {Phys. Rev. D}\ }\textbf {\bibinfo {volume} {96}},\ \bibinfo
  {pages} {085006} (\bibinfo {year} {2017}{\natexlab{b}})},\ \Eprint
  {https://arxiv.org/abs/1706.03917} {arXiv:1706.03917 [hep-th]} \BibitemShut
  {NoStop}%
\bibitem [{\citenamefont {Kapec}\ \emph
  {et~al.}(2017{\natexlab{b}})\citenamefont {Kapec}, \citenamefont {Mitra},
  \citenamefont {Raclariu},\ and\ \citenamefont {Strominger}}]{Kapec:2016jld}%
  \BibitemOpen
  \bibfield  {author} {\bibinfo {author} {\bibfnamefont {D.}~\bibnamefont
  {Kapec}}, \bibinfo {author} {\bibfnamefont {P.}~\bibnamefont {Mitra}},
  \bibinfo {author} {\bibfnamefont {A.-M.}\ \bibnamefont {Raclariu}},\ and\
  \bibinfo {author} {\bibfnamefont {A.}~\bibnamefont {Strominger}},\ }\bibfield
   {title} {\emph {\bibinfo {title} {{2D Stress Tensor for 4D Gravity}}},\
  }\href {https://doi.org/10.1103/PhysRevLett.119.121601} {\bibfield  {journal}
  {\bibinfo  {journal} {Phys. Rev. Lett.}\ }\textbf {\bibinfo {volume} {119}},\
  \bibinfo {pages} {121601} (\bibinfo {year} {2017}{\natexlab{b}})},\ \Eprint
  {https://arxiv.org/abs/1609.00282} {arXiv:1609.00282 [hep-th]} \BibitemShut
  {NoStop}%
\bibitem [{\citenamefont {Nande}\ \emph {et~al.}(2018)\citenamefont {Nande},
  \citenamefont {Pate},\ and\ \citenamefont {Strominger}}]{Nande:2017dba}%
  \BibitemOpen
  \bibfield  {author} {\bibinfo {author} {\bibfnamefont {A.}~\bibnamefont
  {Nande}}, \bibinfo {author} {\bibfnamefont {M.}~\bibnamefont {Pate}},\ and\
  \bibinfo {author} {\bibfnamefont {A.}~\bibnamefont {Strominger}},\ }\bibfield
   {title} {\emph {\bibinfo {title} {{Soft Factorization in QED from 2D
  Kac-Moody Symmetry}}},\ }\href {https://doi.org/10.1007/JHEP02(2018)079}
  {\bibfield  {journal} {\bibinfo  {journal} {JHEP}\ }\textbf {\bibinfo
  {volume} {02}},\ \bibinfo {pages} {079}},\ \Eprint
  {https://arxiv.org/abs/1705.00608} {arXiv:1705.00608 [hep-th]} \BibitemShut
  {NoStop}%
\bibitem [{\citenamefont {Donnay}\ \emph {et~al.}(2019)\citenamefont {Donnay},
  \citenamefont {Puhm},\ and\ \citenamefont {Strominger}}]{Donnay:2018neh}%
  \BibitemOpen
  \bibfield  {author} {\bibinfo {author} {\bibfnamefont {L.}~\bibnamefont
  {Donnay}}, \bibinfo {author} {\bibfnamefont {A.}~\bibnamefont {Puhm}},\ and\
  \bibinfo {author} {\bibfnamefont {A.}~\bibnamefont {Strominger}},\ }\bibfield
   {title} {\emph {\bibinfo {title} {{Conformally Soft Photons and
  Gravitons}}},\ }\href {https://doi.org/10.1007/JHEP01(2019)184} {\bibfield
  {journal} {\bibinfo  {journal} {JHEP}\ }\textbf {\bibinfo {volume} {01}},\
  \bibinfo {pages} {184}},\ \Eprint {https://arxiv.org/abs/1810.05219}
  {arXiv:1810.05219 [hep-th]} \BibitemShut {NoStop}%
\bibitem [{\citenamefont {Fan}\ \emph {et~al.}(2019)\citenamefont {Fan},
  \citenamefont {Fotopoulos},\ and\ \citenamefont {Taylor}}]{Fan:2019emx}%
  \BibitemOpen
  \bibfield  {author} {\bibinfo {author} {\bibfnamefont {W.}~\bibnamefont
  {Fan}}, \bibinfo {author} {\bibfnamefont {A.}~\bibnamefont {Fotopoulos}},\
  and\ \bibinfo {author} {\bibfnamefont {T.~R.}\ \bibnamefont {Taylor}},\
  }\bibfield  {title} {\emph {\bibinfo {title} {{Soft Limits of Yang-Mills
  Amplitudes and Conformal Correlators}}},\ }\href
  {https://doi.org/10.1007/JHEP05(2019)121} {\bibfield  {journal} {\bibinfo
  {journal} {JHEP}\ }\textbf {\bibinfo {volume} {05}},\ \bibinfo {pages}
  {121}},\ \Eprint {https://arxiv.org/abs/1903.01676} {arXiv:1903.01676
  [hep-th]} \BibitemShut {NoStop}%
\bibitem [{\citenamefont {Nandan}\ \emph {et~al.}(2019)\citenamefont {Nandan},
  \citenamefont {Schreiber}, \citenamefont {Volovich},\ and\ \citenamefont
  {Zlotnikov}}]{Nandan:2019jas}%
  \BibitemOpen
  \bibfield  {author} {\bibinfo {author} {\bibfnamefont {D.}~\bibnamefont
  {Nandan}}, \bibinfo {author} {\bibfnamefont {A.}~\bibnamefont {Schreiber}},
  \bibinfo {author} {\bibfnamefont {A.}~\bibnamefont {Volovich}},\ and\
  \bibinfo {author} {\bibfnamefont {M.}~\bibnamefont {Zlotnikov}},\ }\bibfield
  {title} {\emph {\bibinfo {title} {{Celestial Amplitudes: Conformal Partial
  Waves and Soft Limits}}},\ }\href {https://doi.org/10.1007/JHEP10(2019)018}
  {\bibfield  {journal} {\bibinfo  {journal} {JHEP}\ }\textbf {\bibinfo
  {volume} {10}},\ \bibinfo {pages} {018}},\ \Eprint
  {https://arxiv.org/abs/1904.10940} {arXiv:1904.10940 [hep-th]} \BibitemShut
  {NoStop}%
\bibitem [{\citenamefont {Pate}\ \emph {et~al.}(2019)\citenamefont {Pate},
  \citenamefont {Raclariu},\ and\ \citenamefont {Strominger}}]{Pate:2019mfs}%
  \BibitemOpen
  \bibfield  {author} {\bibinfo {author} {\bibfnamefont {M.}~\bibnamefont
  {Pate}}, \bibinfo {author} {\bibfnamefont {A.-M.}\ \bibnamefont {Raclariu}},\
  and\ \bibinfo {author} {\bibfnamefont {A.}~\bibnamefont {Strominger}},\
  }\bibfield  {title} {\emph {\bibinfo {title} {{Conformally Soft Theorem in
  Gauge Theory}}},\ }\href {https://doi.org/10.1103/PhysRevD.100.085017}
  {\bibfield  {journal} {\bibinfo  {journal} {Phys. Rev. D}\ }\textbf {\bibinfo
  {volume} {100}},\ \bibinfo {pages} {085017} (\bibinfo {year} {2019})},\
  \Eprint {https://arxiv.org/abs/1904.10831} {arXiv:1904.10831 [hep-th]}
  \BibitemShut {NoStop}%
\bibitem [{\citenamefont {Adamo}\ \emph {et~al.}(2019)\citenamefont {Adamo},
  \citenamefont {Mason},\ and\ \citenamefont {Sharma}}]{Adamo:2019ipt}%
  \BibitemOpen
  \bibfield  {author} {\bibinfo {author} {\bibfnamefont {T.}~\bibnamefont
  {Adamo}}, \bibinfo {author} {\bibfnamefont {L.}~\bibnamefont {Mason}},\ and\
  \bibinfo {author} {\bibfnamefont {A.}~\bibnamefont {Sharma}},\ }\bibfield
  {title} {\emph {\bibinfo {title} {{Celestial amplitudes and conformal soft
  theorems}}},\ }\href {https://doi.org/10.1088/1361-6382/ab42ce} {\bibfield
  {journal} {\bibinfo  {journal} {Class. Quant. Grav.}\ }\textbf {\bibinfo
  {volume} {36}},\ \bibinfo {pages} {205018} (\bibinfo {year} {2019})},\
  \Eprint {https://arxiv.org/abs/1905.09224} {arXiv:1905.09224 [hep-th]}
  \BibitemShut {NoStop}%
\bibitem [{\citenamefont {Puhm}(2020)}]{Puhm:2019zbl}%
  \BibitemOpen
  \bibfield  {author} {\bibinfo {author} {\bibfnamefont {A.}~\bibnamefont
  {Puhm}},\ }\bibfield  {title} {\emph {\bibinfo {title} {{Conformally Soft
  Theorem in Gravity}}},\ }\href {https://doi.org/10.1007/JHEP09(2020)130}
  {\bibfield  {journal} {\bibinfo  {journal} {JHEP}\ }\textbf {\bibinfo
  {volume} {09}},\ \bibinfo {pages} {130}},\ \Eprint
  {https://arxiv.org/abs/1905.09799} {arXiv:1905.09799 [hep-th]} \BibitemShut
  {NoStop}%
\bibitem [{\citenamefont {Guevara}(2019)}]{Guevara:2019ypd}%
  \BibitemOpen
  \bibfield  {author} {\bibinfo {author} {\bibfnamefont {A.}~\bibnamefont
  {Guevara}},\ }\bibfield  {title} {\emph {\bibinfo {title} {{Notes on
  Conformal Soft Theorems and Recursion Relations in Gravity}}},\ }\href@noop
  {} {\  (\bibinfo {year} {2019})},\ \Eprint {https://arxiv.org/abs/1906.07810}
  {arXiv:1906.07810 [hep-th]} \BibitemShut {NoStop}%
\bibitem [{\citenamefont {Kapec}\ and\ \citenamefont
  {Mitra}(2018)}]{Kapec:2017gsg}%
  \BibitemOpen
  \bibfield  {author} {\bibinfo {author} {\bibfnamefont {D.}~\bibnamefont
  {Kapec}}\ and\ \bibinfo {author} {\bibfnamefont {P.}~\bibnamefont {Mitra}},\
  }\bibfield  {title} {\emph {\bibinfo {title} {{A $d$-Dimensional Stress
  Tensor for Mink$_{d+2}$ Gravity}}},\ }\href
  {https://doi.org/10.1007/JHEP05(2018)186} {\bibfield  {journal} {\bibinfo
  {journal} {JHEP}\ }\textbf {\bibinfo {volume} {05}},\ \bibinfo {pages}
  {186}},\ \Eprint {https://arxiv.org/abs/1711.04371} {arXiv:1711.04371
  [hep-th]} \BibitemShut {NoStop}%
\bibitem [{\citenamefont {Donnay}\ \emph {et~al.}(2020)\citenamefont {Donnay},
  \citenamefont {Pasterski},\ and\ \citenamefont {Puhm}}]{Donnay:2020guq}%
  \BibitemOpen
  \bibfield  {author} {\bibinfo {author} {\bibfnamefont {L.}~\bibnamefont
  {Donnay}}, \bibinfo {author} {\bibfnamefont {S.}~\bibnamefont {Pasterski}},\
  and\ \bibinfo {author} {\bibfnamefont {A.}~\bibnamefont {Puhm}},\ }\bibfield
  {title} {\emph {\bibinfo {title} {{Asymptotic Symmetries and Celestial
  CFT}}},\ }\href {https://doi.org/10.1007/JHEP09(2020)176} {\bibfield
  {journal} {\bibinfo  {journal} {JHEP}\ }\textbf {\bibinfo {volume} {09}},\
  \bibinfo {pages} {176}},\ \Eprint {https://arxiv.org/abs/2005.08990}
  {arXiv:2005.08990 [hep-th]} \BibitemShut {NoStop}%
\bibitem [{\citenamefont {Kapec}\ and\ \citenamefont
  {Mitra}(2022)}]{Kapec:2021eug}%
  \BibitemOpen
  \bibfield  {author} {\bibinfo {author} {\bibfnamefont {D.}~\bibnamefont
  {Kapec}}\ and\ \bibinfo {author} {\bibfnamefont {P.}~\bibnamefont {Mitra}},\
  }\bibfield  {title} {\emph {\bibinfo {title} {{Shadows and soft exchange in
  celestial CFT}}},\ }\href {https://doi.org/10.1103/PhysRevD.105.026009}
  {\bibfield  {journal} {\bibinfo  {journal} {Phys. Rev. D}\ }\textbf {\bibinfo
  {volume} {105}},\ \bibinfo {pages} {026009} (\bibinfo {year} {2022})},\
  \Eprint {https://arxiv.org/abs/2109.00073} {arXiv:2109.00073 [hep-th]}
  \BibitemShut {NoStop}%
\bibitem [{\citenamefont {Pasterski}\ \emph {et~al.}(2021)\citenamefont
  {Pasterski}, \citenamefont {Puhm},\ and\ \citenamefont
  {Trevisani}}]{Pasterski:2021fjn}%
  \BibitemOpen
  \bibfield  {author} {\bibinfo {author} {\bibfnamefont {S.}~\bibnamefont
  {Pasterski}}, \bibinfo {author} {\bibfnamefont {A.}~\bibnamefont {Puhm}},\
  and\ \bibinfo {author} {\bibfnamefont {E.}~\bibnamefont {Trevisani}},\
  }\bibfield  {title} {\emph {\bibinfo {title} {{Celestial Diamonds: Conformal
  Multiplets in Celestial CFT}}},\ }\href@noop {} {\  (\bibinfo {year}
  {2021})},\ \Eprint {https://arxiv.org/abs/2105.03516} {arXiv:2105.03516
  [hep-th]} \BibitemShut {NoStop}%
\bibitem [{\citenamefont {Donnay}\ \emph
  {et~al.}(2022{\natexlab{a}})\citenamefont {Donnay}, \citenamefont
  {Pasterski},\ and\ \citenamefont {Puhm}}]{Donnay:2022sdg}%
  \BibitemOpen
  \bibfield  {author} {\bibinfo {author} {\bibfnamefont {L.}~\bibnamefont
  {Donnay}}, \bibinfo {author} {\bibfnamefont {S.}~\bibnamefont {Pasterski}},\
  and\ \bibinfo {author} {\bibfnamefont {A.}~\bibnamefont {Puhm}},\ }\bibfield
  {title} {\emph {\bibinfo {title} {{Goldilocks modes and the three scattering
  bases}}},\ }\href {https://doi.org/10.1007/JHEP06(2022)124} {\bibfield
  {journal} {\bibinfo  {journal} {JHEP}\ }\textbf {\bibinfo {volume} {06}},\
  \bibinfo {pages} {124}},\ \Eprint {https://arxiv.org/abs/2202.11127}
  {arXiv:2202.11127 [hep-th]} \BibitemShut {NoStop}%
\bibitem [{\citenamefont {Pano}\ \emph {et~al.}(2023)\citenamefont {Pano},
  \citenamefont {Puhm},\ and\ \citenamefont {Trevisani}}]{Pano:2023slc}%
  \BibitemOpen
  \bibfield  {author} {\bibinfo {author} {\bibfnamefont {Y.}~\bibnamefont
  {Pano}}, \bibinfo {author} {\bibfnamefont {A.}~\bibnamefont {Puhm}},\ and\
  \bibinfo {author} {\bibfnamefont {E.}~\bibnamefont {Trevisani}},\ }\bibfield
  {title} {\emph {\bibinfo {title} {{Symmetries in Celestial CFT$_{d}$}}},\
  }\href {https://doi.org/10.1007/JHEP07(2023)076} {\bibfield  {journal}
  {\bibinfo  {journal} {JHEP}\ }\textbf {\bibinfo {volume} {07}},\ \bibinfo
  {pages} {076}},\ \Eprint {https://arxiv.org/abs/2302.10222} {arXiv:2302.10222
  [hep-th]} \BibitemShut {NoStop}%
\bibitem [{\citenamefont {Guevara}\ \emph {et~al.}(2021)\citenamefont
  {Guevara}, \citenamefont {Himwich}, \citenamefont {Pate},\ and\ \citenamefont
  {Strominger}}]{Guevara:2021abz}%
  \BibitemOpen
  \bibfield  {author} {\bibinfo {author} {\bibfnamefont {A.}~\bibnamefont
  {Guevara}}, \bibinfo {author} {\bibfnamefont {E.}~\bibnamefont {Himwich}},
  \bibinfo {author} {\bibfnamefont {M.}~\bibnamefont {Pate}},\ and\ \bibinfo
  {author} {\bibfnamefont {A.}~\bibnamefont {Strominger}},\ }\bibfield  {title}
  {\emph {\bibinfo {title} {{Holographic Symmetry Algebras for Gauge Theory and
  Gravity}}},\ }\href@noop {} {\  (\bibinfo {year} {2021})},\ \Eprint
  {https://arxiv.org/abs/2103.03961} {arXiv:2103.03961 [hep-th]} \BibitemShut
  {NoStop}%
\bibitem [{\citenamefont {Strominger}(2021)}]{Strominger:2021mtt}%
  \BibitemOpen
  \bibfield  {author} {\bibinfo {author} {\bibfnamefont {A.}~\bibnamefont
  {Strominger}},\ }\bibfield  {title} {\emph {\bibinfo {title} {{$w_{1+\infty}$
  Algebra and the Celestial Sphere: Infinite Towers of Soft Graviton, Photon,
  and Gluon Symmetries}}},\ }\href
  {https://doi.org/10.1103/PhysRevLett.127.221601} {\bibfield  {journal}
  {\bibinfo  {journal} {Phys. Rev. Lett.}\ }\textbf {\bibinfo {volume} {127}},\
  \bibinfo {pages} {221601} (\bibinfo {year} {2021})},\ \Eprint
  {https://arxiv.org/abs/2105.14346} {arXiv:2105.14346 [hep-th]} \BibitemShut
  {NoStop}%
\bibitem [{\citenamefont {Geiller}(2024)}]{Geiller:2024bgf}%
  \BibitemOpen
  \bibfield  {author} {\bibinfo {author} {\bibfnamefont {M.}~\bibnamefont
  {Geiller}},\ }\bibfield  {title} {\emph {\bibinfo {title} {{Celestial
  $w_{1+\infty}$ charges and the subleading structure of asymptotically-flat
  spacetimes}}},\ }\href@noop {} {\  (\bibinfo {year} {2024})},\ \Eprint
  {https://arxiv.org/abs/2403.05195} {arXiv:2403.05195 [hep-th]} \BibitemShut
  {NoStop}%
\bibitem [{\citenamefont {Penrose}(1976)}]{Penrose:1976js}%
  \BibitemOpen
  \bibfield  {author} {\bibinfo {author} {\bibfnamefont {R.}~\bibnamefont
  {Penrose}},\ }\bibfield  {title} {\emph {\bibinfo {title} {{Nonlinear
  Gravitons and Curved Twistor Theory}}},\ }\href
  {https://doi.org/10.1007/BF00762011} {\bibfield  {journal} {\bibinfo
  {journal} {Gen. Rel. Grav.}\ }\textbf {\bibinfo {volume} {7}},\ \bibinfo
  {pages} {31} (\bibinfo {year} {1976})}\BibitemShut {NoStop}%
\bibitem [{Note2()}]{Note2}%
  \BibitemOpen
  \bibinfo {note} {Note that in $D>4$ dimensions these soft theorems are
  all-loop exact.}\BibitemShut {Stop}%
\bibitem [{\citenamefont {Bern}\ \emph {et~al.}(2014)\citenamefont {Bern},
  \citenamefont {Davies},\ and\ \citenamefont {Nohle}}]{Bern:2014oka}%
  \BibitemOpen
  \bibfield  {author} {\bibinfo {author} {\bibfnamefont {Z.}~\bibnamefont
  {Bern}}, \bibinfo {author} {\bibfnamefont {S.}~\bibnamefont {Davies}},\ and\
  \bibinfo {author} {\bibfnamefont {J.}~\bibnamefont {Nohle}},\ }\bibfield
  {title} {\emph {\bibinfo {title} {{On Loop Corrections to Subleading Soft
  Behavior of Gluons and Gravitons}}},\ }\href
  {https://doi.org/10.1103/PhysRevD.90.085015} {\bibfield  {journal} {\bibinfo
  {journal} {Phys. Rev. D}\ }\textbf {\bibinfo {volume} {90}},\ \bibinfo
  {pages} {085015} (\bibinfo {year} {2014})},\ \Eprint
  {https://arxiv.org/abs/1405.1015} {arXiv:1405.1015 [hep-th]} \BibitemShut
  {NoStop}%
\bibitem [{\citenamefont {Sahoo}\ and\ \citenamefont
  {Sen}(2019)}]{Sahoo:2018lxl}%
  \BibitemOpen
  \bibfield  {author} {\bibinfo {author} {\bibfnamefont {B.}~\bibnamefont
  {Sahoo}}\ and\ \bibinfo {author} {\bibfnamefont {A.}~\bibnamefont {Sen}},\
  }\bibfield  {title} {\emph {\bibinfo {title} {{Classical and Quantum Results
  on Logarithmic Terms in the Soft Theorem in Four Dimensions}}},\ }\href
  {https://doi.org/10.1007/JHEP02(2019)086} {\bibfield  {journal} {\bibinfo
  {journal} {JHEP}\ }\textbf {\bibinfo {volume} {02}},\ \bibinfo {pages}
  {086}},\ \Eprint {https://arxiv.org/abs/1808.03288} {arXiv:1808.03288
  [hep-th]} \BibitemShut {NoStop}%
\bibitem [{\citenamefont {Saha}\ \emph {et~al.}(2020)\citenamefont {Saha},
  \citenamefont {Sahoo},\ and\ \citenamefont {Sen}}]{Saha:2019tub}%
  \BibitemOpen
  \bibfield  {author} {\bibinfo {author} {\bibfnamefont {A.~P.}\ \bibnamefont
  {Saha}}, \bibinfo {author} {\bibfnamefont {B.}~\bibnamefont {Sahoo}},\ and\
  \bibinfo {author} {\bibfnamefont {A.}~\bibnamefont {Sen}},\ }\bibfield
  {title} {\emph {\bibinfo {title} {{Proof of the classical soft graviton
  theorem in $D$ = 4}}},\ }\href {https://doi.org/10.1007/JHEP06(2020)153}
  {\bibfield  {journal} {\bibinfo  {journal} {JHEP}\ }\textbf {\bibinfo
  {volume} {06}},\ \bibinfo {pages} {153}},\ \Eprint
  {https://arxiv.org/abs/1912.06413} {arXiv:1912.06413 [hep-th]} \BibitemShut
  {NoStop}%
\bibitem [{\citenamefont {Sahoo}(2020)}]{Sahoo:2020ryf}%
  \BibitemOpen
  \bibfield  {author} {\bibinfo {author} {\bibfnamefont {B.}~\bibnamefont
  {Sahoo}},\ }\bibfield  {title} {\emph {\bibinfo {title} {{Classical
  Sub-subleading Soft Photon and Soft Graviton Theorems in Four Spacetime
  Dimensions}}},\ }\href {https://doi.org/10.1007/JHEP12(2020)070} {\bibfield
  {journal} {\bibinfo  {journal} {JHEP}\ }\textbf {\bibinfo {volume} {12}},\
  \bibinfo {pages} {070}},\ \Eprint {https://arxiv.org/abs/2008.04376}
  {arXiv:2008.04376 [hep-th]} \BibitemShut {NoStop}%
\bibitem [{\citenamefont {Campiglia}\ and\ \citenamefont
  {Laddha}(2019{\natexlab{b}})}]{Campiglia:2019wxe}%
  \BibitemOpen
  \bibfield  {author} {\bibinfo {author} {\bibfnamefont {M.}~\bibnamefont
  {Campiglia}}\ and\ \bibinfo {author} {\bibfnamefont {A.}~\bibnamefont
  {Laddha}},\ }\bibfield  {title} {\emph {\bibinfo {title} {{Loop Corrected
  Soft Photon Theorem as a Ward Identity}}},\ }\href
  {https://doi.org/10.1007/JHEP10(2019)287} {\bibfield  {journal} {\bibinfo
  {journal} {JHEP}\ }\textbf {\bibinfo {volume} {10}},\ \bibinfo {pages}
  {287}},\ \Eprint {https://arxiv.org/abs/1903.09133} {arXiv:1903.09133
  [hep-th]} \BibitemShut {NoStop}%
\bibitem [{\citenamefont {Atul~Bhatkar}(2020)}]{AtulBhatkar:2019vcb}%
  \BibitemOpen
  \bibfield  {author} {\bibinfo {author} {\bibfnamefont {S.}~\bibnamefont
  {Atul~Bhatkar}},\ }\bibfield  {title} {\emph {\bibinfo {title} {{Ward
  identity for loop level soft photon theorem for massless QED coupled to
  gravity}}},\ }\href {https://doi.org/10.1007/JHEP10(2020)110} {\bibfield
  {journal} {\bibinfo  {journal} {JHEP}\ }\textbf {\bibinfo {volume} {10}},\
  \bibinfo {pages} {110}},\ \Eprint {https://arxiv.org/abs/1912.10229}
  {arXiv:1912.10229 [hep-th]} \BibitemShut {NoStop}%
\bibitem [{Note3()}]{Note3}%
  \BibitemOpen
  \bibinfo {note} {We thank E.~Trevisani for discussion on this
  point.}\BibitemShut {Stop}%
\bibitem [{Note4()}]{Note4}%
  \BibitemOpen
  \bibinfo {note} {Related recent work includes \cite
  {Krishna:2023ukw,Bhardwaj:2024wld}}\BibitemShut {NoStop}%
\bibitem [{\citenamefont {Donnay}\ \emph
  {et~al.}(2022{\natexlab{b}})\citenamefont {Donnay}, \citenamefont {Nguyen},\
  and\ \citenamefont {Ruzziconi}}]{Donnay:2022hkf}%
  \BibitemOpen
  \bibfield  {author} {\bibinfo {author} {\bibfnamefont {L.}~\bibnamefont
  {Donnay}}, \bibinfo {author} {\bibfnamefont {K.}~\bibnamefont {Nguyen}},\
  and\ \bibinfo {author} {\bibfnamefont {R.}~\bibnamefont {Ruzziconi}},\
  }\bibfield  {title} {\emph {\bibinfo {title} {{Loop-corrected subleading soft
  theorem and the celestial stress tensor}}},\ }\href
  {https://doi.org/10.1007/JHEP09(2022)063} {\bibfield  {journal} {\bibinfo
  {journal} {JHEP}\ }\textbf {\bibinfo {volume} {09}},\ \bibinfo {pages}
  {063}},\ \Eprint {https://arxiv.org/abs/2205.11477} {arXiv:2205.11477
  [hep-th]} \BibitemShut {NoStop}%
\bibitem [{\citenamefont {Agrawal}\ \emph {et~al.}(2024)\citenamefont
  {Agrawal}, \citenamefont {Donnay}, \citenamefont {Nguyen},\ and\
  \citenamefont {Ruzziconi}}]{Agrawal:2023zea}%
  \BibitemOpen
  \bibfield  {author} {\bibinfo {author} {\bibfnamefont {S.}~\bibnamefont
  {Agrawal}}, \bibinfo {author} {\bibfnamefont {L.}~\bibnamefont {Donnay}},
  \bibinfo {author} {\bibfnamefont {K.}~\bibnamefont {Nguyen}},\ and\ \bibinfo
  {author} {\bibfnamefont {R.}~\bibnamefont {Ruzziconi}},\ }\bibfield  {title}
  {\emph {\bibinfo {title} {{Logarithmic soft graviton theorems from
  superrotation Ward identities}}},\ }\href
  {https://doi.org/10.1007/JHEP02(2024)120} {\bibfield  {journal} {\bibinfo
  {journal} {JHEP}\ }\textbf {\bibinfo {volume} {02}},\ \bibinfo {pages}
  {120}},\ \Eprint {https://arxiv.org/abs/2309.11220} {arXiv:2309.11220
  [hep-th]} \BibitemShut {NoStop}%
\bibitem [{\citenamefont {Choi}\ \emph
  {et~al.}(2024{\natexlab{a}})\citenamefont {Choi}, \citenamefont {Laddha},\
  and\ \citenamefont {Puhm}}]{Choi:2024ajz}%
  \BibitemOpen
  \bibfield  {author} {\bibinfo {author} {\bibfnamefont {S.}~\bibnamefont
  {Choi}}, \bibinfo {author} {\bibfnamefont {A.}~\bibnamefont {Laddha}},\ and\
  \bibinfo {author} {\bibfnamefont {A.}~\bibnamefont {Puhm}},\ }\bibfield
  {title} {\emph {\bibinfo {title} {{The Classical Super-Rotation Infrared
  Triangle}}},\ }\href@noop {} {\  (\bibinfo {year} {2024}{\natexlab{a}})},\
  \Eprint {https://arxiv.org/abs/2412.16142} {arXiv:2412.16142 [hep-th]}
  \BibitemShut {NoStop}%
\bibitem [{\citenamefont {Choi}\ \emph
  {et~al.}(2024{\natexlab{b}})\citenamefont {Choi}, \citenamefont {Laddha},\
  and\ \citenamefont {Puhm}}]{Choi:2024mac}%
  \BibitemOpen
  \bibfield  {author} {\bibinfo {author} {\bibfnamefont {S.}~\bibnamefont
  {Choi}}, \bibinfo {author} {\bibfnamefont {A.}~\bibnamefont {Laddha}},\ and\
  \bibinfo {author} {\bibfnamefont {A.}~\bibnamefont {Puhm}},\ }\bibfield
  {title} {\emph {\bibinfo {title} {{The Classical Super-Phaserotation Infrared
  Triangle}}},\ }\href@noop {} {\  (\bibinfo {year} {2024}{\natexlab{b}})},\
  \Eprint {https://arxiv.org/abs/2412.16149} {arXiv:2412.16149 [hep-th]}
  \BibitemShut {NoStop}%
\bibitem [{\citenamefont {Campiglia}(2015)}]{Campiglia:2015lxa}%
  \BibitemOpen
  \bibfield  {author} {\bibinfo {author} {\bibfnamefont {M.}~\bibnamefont
  {Campiglia}},\ }\bibfield  {title} {\emph {\bibinfo {title} {{Null to
  time-like infinity Green\textquoteright{}s functions for asymptotic
  symmetries in Minkowski spacetime}}},\ }\href
  {https://doi.org/10.1007/JHEP11(2015)160} {\bibfield  {journal} {\bibinfo
  {journal} {JHEP}\ }\textbf {\bibinfo {volume} {11}},\ \bibinfo {pages}
  {160}},\ \Eprint {https://arxiv.org/abs/1509.01408} {arXiv:1509.01408
  [hep-th]} \BibitemShut {NoStop}%
\bibitem [{\citenamefont {Peraza}(2024)}]{Peraza:2023ivy}%
  \BibitemOpen
  \bibfield  {author} {\bibinfo {author} {\bibfnamefont {J.}~\bibnamefont
  {Peraza}},\ }\bibfield  {title} {\emph {\bibinfo {title} {{Renormalized
  electric and magnetic charges for O(r$^{n}$) large gauge symmetries}}},\
  }\href {https://doi.org/10.1007/JHEP01(2024)175} {\bibfield  {journal}
  {\bibinfo  {journal} {JHEP}\ }\textbf {\bibinfo {volume} {01}},\ \bibinfo
  {pages} {175}},\ \Eprint {https://arxiv.org/abs/2301.05671} {arXiv:2301.05671
  [hep-th]} \BibitemShut {NoStop}%
\bibitem [{Note5()}]{Note5}%
  \BibitemOpen
  \bibinfo {note} {In the literature, LGTs with divergent (``overleading'')
  gauge parameters are sometimes referred to as overleading LGTs; here we
  instead we refer to them as subleading LGTs since they are related to
  subleading soft theorems.}\BibitemShut {Stop}%
\bibitem [{\citenamefont {Christodoulou}( DOI)}]{GIVPCD}%
  \BibitemOpen
  \bibfield  {author} {\bibinfo {author} {\bibfnamefont {D.}~\bibnamefont
  {Christodoulou}},\ }\bibfield  {title} {\emph {\bibinfo {title} {The global
  initial value problem in general relativity}},\ }\href@noop {} {\bibfield
  {journal} {\bibinfo  {journal} {Proceedings of the MGIX MM Meeting. Held 2-8
  July 2000 in The University of Rome "La Sapienza",}\ } (\bibinfo {year} {Pub
  Date: December 2002 DOI:})}\BibitemShut {NoStop}%
\bibitem [{\citenamefont {Damour}(1986)}]{tdamour1986}%
  \BibitemOpen
  \bibfield  {author} {\bibinfo {author} {\bibfnamefont {T.}~\bibnamefont
  {Damour}},\ }\bibfield  {title} {\emph {\bibinfo {title} {Analytical
  calculations of gravitational radiation.}},\ }\href@noop {} {\bibfield
  {journal} {\bibinfo  {journal} {In: Proceedings of the Fourth Marcel
  Grossmann Meeting on General Relativity, , Elsevier Science Publishers}\ ,\
  \bibinfo {pages} {pp. 365}} (\bibinfo {year} {1986})}\BibitemShut {NoStop}%
\bibitem [{\citenamefont {Kehrberger}(2022)}]{Kehrberger:2021uvf}%
  \BibitemOpen
  \bibfield  {author} {\bibinfo {author} {\bibfnamefont {L.~M.~A.}\
  \bibnamefont {Kehrberger}},\ }\bibfield  {title} {\emph {\bibinfo {title}
  {{The Case Against Smooth Null Infinity I: Heuristics and
  Counter-Examples}}},\ }\href {https://doi.org/10.1007/s00023-021-01108-2}
  {\bibfield  {journal} {\bibinfo  {journal} {Annales Henri Poincare}\ }\textbf
  {\bibinfo {volume} {23}},\ \bibinfo {pages} {829} (\bibinfo {year} {2022})},\
  \Eprint {https://arxiv.org/abs/2105.08079} {arXiv:2105.08079 [gr-qc]}
  \BibitemShut {NoStop}%
\bibitem [{\citenamefont {Geiller}\ and\ \citenamefont
  {Zwikel}(2022)}]{Geiller:2022vto}%
  \BibitemOpen
  \bibfield  {author} {\bibinfo {author} {\bibfnamefont {M.}~\bibnamefont
  {Geiller}}\ and\ \bibinfo {author} {\bibfnamefont {C.}~\bibnamefont
  {Zwikel}},\ }\bibfield  {title} {\emph {\bibinfo {title} {{The partial Bondi
  gauge: Further enlarging the asymptotic structure of gravity}}},\ }\href@noop
  {} {\bibfield  {journal} {\bibinfo  {journal} {SciPost Phys.}\ }\textbf
  {\bibinfo {volume} {13}},\ \bibinfo {pages} {108} (\bibinfo {year}
  {2022})}\BibitemShut {NoStop}%
\bibitem [{\citenamefont {Geiller}\ and\ \citenamefont
  {Zwikel}(2024)}]{Geiller:2024amx}%
  \BibitemOpen
  \bibfield  {author} {\bibinfo {author} {\bibfnamefont {M.}~\bibnamefont
  {Geiller}}\ and\ \bibinfo {author} {\bibfnamefont {C.}~\bibnamefont
  {Zwikel}},\ }\bibfield  {title} {\emph {\bibinfo {title} {{The partial Bondi
  gauge: Gauge fixings and asymptotic charges}}},\ }\href@noop {} {\  (\bibinfo
  {year} {2024})}\BibitemShut {NoStop}%
\bibitem [{\citenamefont {Geiller}\ \emph {et~al.}(2024)\citenamefont
  {Geiller}, \citenamefont {Laddha},\ and\ \citenamefont
  {Zwikel}}]{Geiller:2024ryw}%
  \BibitemOpen
  \bibfield  {author} {\bibinfo {author} {\bibfnamefont {M.}~\bibnamefont
  {Geiller}}, \bibinfo {author} {\bibfnamefont {A.}~\bibnamefont {Laddha}},\
  and\ \bibinfo {author} {\bibfnamefont {C.}~\bibnamefont {Zwikel}},\
  }\bibfield  {title} {\emph {\bibinfo {title} {{Symmetries of the
  gravitational scattering in the absence of peeling}}},\ }\href
  {https://doi.org/10.1007/JHEP12(2024)081} {\bibfield  {journal} {\bibinfo
  {journal} {JHEP}\ }\textbf {\bibinfo {volume} {12}},\ \bibinfo {pages}
  {081}},\ \Eprint {https://arxiv.org/abs/2407.07978} {arXiv:2407.07978
  [gr-qc]} \BibitemShut {NoStop}%
\bibitem [{Note6()}]{Note6}%
  \BibitemOpen
  \bibinfo {note} {As the gravitational soft theorems have been derived in
  harmonic gauge, we believe that the logarithmic divergence of the
  superrotation charge is a physical effect referred to as tail to the memory
  in \cite {Laddha:2018vbn}.}\BibitemShut {Stop}%
\bibitem [{\citenamefont {Strominger}(2017)}]{Strominger:2017zoo}%
  \BibitemOpen
  \bibfield  {author} {\bibinfo {author} {\bibfnamefont {A.}~\bibnamefont
  {Strominger}},\ }\bibfield  {title} {\emph {\bibinfo {title} {{Lectures on
  the Infrared Structure of Gravity and Gauge Theory}}},\ }\href@noop {} {\
  (\bibinfo {year} {2017})},\ \Eprint {https://arxiv.org/abs/1703.05448}
  {arXiv:1703.05448 [hep-th]} \BibitemShut {NoStop}%
\bibitem [{\citenamefont {Laddha}\ and\ \citenamefont
  {Sen}(2019)}]{Laddha:2018vbn}%
  \BibitemOpen
  \bibfield  {author} {\bibinfo {author} {\bibfnamefont {A.}~\bibnamefont
  {Laddha}}\ and\ \bibinfo {author} {\bibfnamefont {A.}~\bibnamefont {Sen}},\
  }\bibfield  {title} {\emph {\bibinfo {title} {{Observational Signature of the
  Logarithmic Terms in the Soft Graviton Theorem}}},\ }\href
  {https://doi.org/10.1103/PhysRevD.100.024009} {\bibfield  {journal} {\bibinfo
   {journal} {Phys. Rev. D}\ }\textbf {\bibinfo {volume} {100}},\ \bibinfo
  {pages} {024009} (\bibinfo {year} {2019})},\ \Eprint
  {https://arxiv.org/abs/1806.01872} {arXiv:1806.01872 [hep-th]} \BibitemShut
  {NoStop}%
\bibitem [{Note7()}]{Note7}%
  \BibitemOpen
  \bibinfo {note} {There is an additional drag term that has to be treated
  separately, as we will explain in \cite {Choi:2024ajz}.}\BibitemShut {Stop}%
\bibitem [{\citenamefont {Choi}\ \emph {et~al.}(ress)\citenamefont {Choi},
  \citenamefont {Laddha},\ and\ \citenamefont {Puhm}}]{elsewhere}%
  \BibitemOpen
  \bibfield  {author} {\bibinfo {author} {\bibfnamefont {S.}~\bibnamefont
  {Choi}}, \bibinfo {author} {\bibfnamefont {A.}~\bibnamefont {Laddha}},\ and\
  \bibinfo {author} {\bibfnamefont {A.}~\bibnamefont {Puhm}},\ }\href@noop {}
  {\  (\bibinfo {year} {work in progress})}\BibitemShut {NoStop}%
\bibitem [{\citenamefont {Krishna}(2023)}]{Krishna:2023ukw}%
  \BibitemOpen
  \bibfield  {author} {\bibinfo {author} {\bibfnamefont {H.}~\bibnamefont
  {Krishna}},\ }\bibfield  {title} {\emph {\bibinfo {title} {{Celestial gluon
  and graviton OPE at loop level}}},\ }\href@noop {} {\  (\bibinfo {year}
  {2023})},\ \Eprint {https://arxiv.org/abs/2310.16687} {arXiv:2310.16687
  [hep-th]} \BibitemShut {NoStop}%
\bibitem [{\citenamefont {Bhardwaj}\ and\ \citenamefont
  {Yelleshpur~Srikant}(2024)}]{Bhardwaj:2024wld}%
  \BibitemOpen
  \bibfield  {author} {\bibinfo {author} {\bibfnamefont {R.}~\bibnamefont
  {Bhardwaj}}\ and\ \bibinfo {author} {\bibfnamefont {A.}~\bibnamefont
  {Yelleshpur~Srikant}},\ }\bibfield  {title} {\emph {\bibinfo {title}
  {{Celestial soft currents at one-loop and their OPEs}}},\ }\href@noop {} {\
  (\bibinfo {year} {2024})},\ \Eprint {https://arxiv.org/abs/2403.10443}
  {arXiv:2403.10443 [hep-th]} \BibitemShut {NoStop}%
\end{thebibliography}%

\end{document}